%% file: main.tex
\documentclass[11pt]{article}

\usepackage[preprint]{acl}

\usepackage{times}
\usepackage{latexsym}

\usepackage[T1]{fontenc}
\usepackage[utf8]{inputenc}

\usepackage{microtype}

\usepackage{inconsolata}

\usepackage{graphicx}
\usepackage{booktabs}
\usepackage{tabularx}
\usepackage{multirow}
\usepackage{amsmath}
\usepackage{xcolor}
\usepackage{caption}
\usepackage{placeins}

\renewcommand{\floatpagefraction}{0.85}

\renewcommand{\dblfloatpagefraction}{0.85}
\usepackage{tikz}
\usetikzlibrary{arrows.meta, positioning, calc}

\title{Compact-Bench: Measuring What Context Compaction Costs\\Long-Horizon Coding Agents}

\author{
  Ruopei Sun\thanks{Equal contribution.} \quad Jianfeng Cai\footnotemark[1] \quad Kangwen Zhao \quad
  Wengang Zhou \quad Houqiang Li \\
  University of Science and Technology of China
}
\hypersetup{
  pdftitle={Compact-Bench: Measuring What Context Compaction Costs Long-Horizon Coding Agents},
  pdfauthor={Ruopei Sun, Jianfeng Cai, Kangwen Zhao, Wengang Zhou, Houqiang Li},
  pdfsubject={Context compaction for long-horizon coding agents}
}

\begin{document}
\maketitle

\begin{abstract}
Context compaction lets long-horizon coding agents continue beyond their context windows, but can erase the state needed to finish a task. Prior evaluations measure efficiency, retained information, or end-to-end success, leaving the effect of a specific history replacement difficult to isolate. \textbf{Compact-Bench} pairs full-history and summary-only continuations from a shared repository state, and evaluates recurring details and successive production compactions. It covers $35$ SWE-bench repairs, $146$ agentic sessions, and $106$ production chains. Across $14$ summarizer backbones, stable damage ranges from $5-64\%$ of $22$ full-history successes: Claude Opus~$4.8$ preserves the most repairs, while DeepSeek-Flash damages the most. With Kimi-K3 as summarizer, the Kimi Code prompt preserves the most task continuations ($19/22$), whereas Claude Code retains the most recurring details ($15.0\%$). LLM summaries retain at most $16.0\%$ of recurring details, compared with $79.1\%$ for mechanical extraction; production summaries retain $15.7\%$ at the first boundary. Failures recur across methods: one repair is damaged by $13/14$ backbones, and another by all $5$ prompts. These results show that reliable compaction requires preserving the progress needed to complete a task, and that retaining relevant information alone does not ensure successful continuation. \href{https://github.com/cai-jianfeng/compact-bench}{Code} and \href{https://huggingface.co/datasets/caijanfeng/compact-bench}{data} are available online.
\end{abstract}

\section{Introduction}
\label{sec:intro}

Long-horizon coding agents accumulate instructions, tool observations, test results, and partial decisions. Compaction allows them to work beyond the context window, but the resulting summary must carry enough state to resume unfinished work. Long-context benchmarks test retrieval, reasoning, and generation from extended inputs \citep{longbench, infinitebench, ruler, zeroscrolls, leval, loft, longbenchv2, babilong, longgenbench}, and information access can depend on input position \citep{lostinmiddle}. Agent histories add a further requirement: preserving which actions have been taken, which hypotheses have been rejected, and what remains to be done.

End-to-end benchmarks evaluate repository repair, code generation, and interactive tool use \citep{swebench, sweagent, openhands, agentless, livecodebench, agentbench, taubench, webarena, osworld}. A final score, however, does not isolate the effect of replacing a history midway through a task. Recent compaction evaluations examine constraint loss, archived-fact retrieval, future-trajectory support, and execution instability; methods also learn or schedule summarization and remove observations \citep{compint, lantern, slipstream, trace, selfcompact, resum, contextfolding, compactionrl, parallelcompaction, complexitytrap}. Compact-Bench targets the history replacement itself: it compares full-history and summary-only continuation at a shared boundary, then examines detail retention in real agent sessions.

The benchmark asks: \emph{For the same task and pre-compaction state, how does continuation change when an agent receives only a summary instead of its full history?} \textbf{Task Preservation} compares full-history and summary-only continuations from the same repository snapshot, using the same agent, step budget, and executable tests. \textbf{Recurring-Detail Retention} measures preservation of structured details that recur in later requests across $146$ sessions ($143$ real and $3$ simulator-synthesized). \textbf{Production-Chain Retention} follows these details across successive compactions in $106$ real sessions. Together, the three evaluations connect controlled task outcomes with information retention in deployed workflows.

Across $14$ summarizer backbones, stable Task Preservation damage ranges from $5-64\%$ of the $22$ validated full-history successes, with Claude Opus~$4.8$ preserving the most repairs. Of $126$ instance-backbone pairs damaged in the first round, $120$ meet the three-run confirmation criterion. Task preservation and detail retention also distinguish different configurations: Kimi Code gives the most passing continuations ($19/22$), whereas Claude Code retains the most recurring details ($15.0\%$). Paired traces reveal a shared failure pattern: summaries can retain the target file and relevant strings while losing the progress needed to finish a repair. Overall, our contributions are:
\begin{itemize}
    \item We introduce \textbf{Compact-Bench}, a benchmark for evaluating whether context compaction preserves a coding agent's ability to resume work. It combines controlled task continuations with recurring-detail and production-chain measurements.
    \item We compare $14$ summarizer backbones and $5$ production prompts. Stable task damage ranges from $5-64\%$ among $22$ full-history successes, with Claude Opus~$4.8$ preserving the most repairs. With Kimi-K3 as summarizer, Kimi Code preserves the most task continuations ($19/22$), while Claude Code retains the most recurring details ($15.0\%$).
    \item We show that retained repair evidence does not guarantee successful continuation. Paired traces reveal failures even when summaries preserve target files and relevant code strings. Together with the different task-preservation and detail-retention leaders, this finding motivates evaluating compaction through both retained information and agent behavior.
\end{itemize}

\begin{figure*}[t]
\centering
\resizebox{\textwidth}{!}{%
\includegraphics[trim=2bp 2bp 2bp 2bp,clip,width=17.8cm,height=5.5cm]{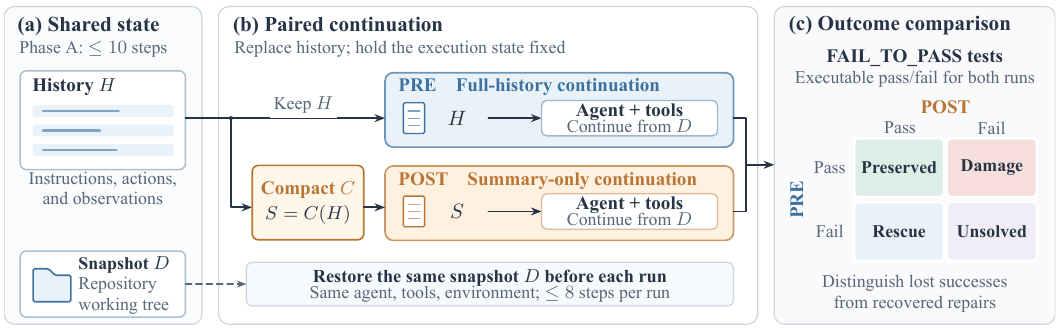}}%
\caption{Paired Task Preservation replay. (a) Phase A fixes history $H$ and repository snapshot $D$. (b) PRE receives $H$ directly; only POST receives the summary $S=C(H)$. Each run restores $D$ and uses matched agent, tool, environment, and step-budget conditions. (c) Repository FAIL\_TO\_PASS tests classify the paired outcomes; rows denote PRE and columns denote POST.}
\label{fig:protocol}
\end{figure*}

\section{Related Work}
\label{sec:related}

\paragraph{Direct evaluations of compaction.}
COMPINT evaluates side-constraint retention, LANTERN tests archived-conversation retrieval, and Slipstream checks summary support for subsequent trajectory steps \citep{compint, lantern, slipstream}. TRACE compares execution after restoring a shared environment, while Complexity Trap evaluates observation masking \citep{trace, complexitytrap}. Compact-Bench combines fixed-boundary replay with a fourteen-backbone and five-prompt comparison, recurring-detail measurements, and repeated production boundaries. The replay links a specific history replacement to a change in task outcome.

\paragraph{Methods for agent histories and persistent memory.}
ReSum summarizes search trajectories; Context-Folding condenses completed subtasks; CompactionRL trains execution and summarization jointly; SelfCompact lets the agent choose when to compact; and parallel compaction summarizes separate blocks \citep{resum, contextfolding, compactionrl, selfcompact, parallelcompaction}. Persistent-memory methods maintain or retrieve information outside the active history, including MemoryBank, A-Mem, MemGPT, and retrieval-augmented generation \citep{memorybank, amem, memgpt, rag}. LongMemEval, LoCoMo, and Beyond Prompts evaluate memory over extended or interleaved dialogue \citep{longmemeval, locomo, beyondprompts}. Our recurring-detail and production-chain tracks assess structured information that reappears after compaction in coding-agent workflows.

\paragraph{Prompt compression and summary evaluation.}
Prompt compression reduces input length through token selection (Selective Context, LLMLingua), question-aware selection (LongLLMLingua), distilled selection (LLMLingua-2), soft representations (Gist tokens, ICAE), and offline compression with adaptation (LLoCO) \citep{selectivecontext, llmlingua, longllmlingua, llmlingua2, gist, icae, lloco}. RECOMP compresses retrieved passages; a survey organizes the broader method families \citep{recomp, promptcompsurvey}. Summary evaluation measures factual consistency and answerability, including Kryscinski et al., FEQA, and QuestEval; G-Eval uses LLM judgments for broader generation quality \citep{kryscinski, feqa, questeval, geval}. Compact-Bench extends the evaluation target to resuming agent work, combining executable outcomes with deterministic matching of recurring details.
\section{Compact-Bench}
\label{sec:benchmark}

\subsection{Evaluation Dimensions}
\label{sec:protocol}

Let $H$ denote the serialized agent history immediately before a compaction boundary. The compaction method under test, denoted by $C$, includes the summarizer and prompt together with the fixed serialization and truncation procedure described in Appendix~\ref{app:harness}; applying it to $H$ produces summary $S=C(H)$. A compaction boundary is the point at which this replacement occurs. In Task Preservation, PRE continues with the full history $H$, whereas POST continues with only $S$. Both conditions start from the same repository working-tree snapshot, denoted by $D$ in Figure~\ref{fig:protocol}.

\textbf{Task Preservation} tests whether continuation from $S$ completes repairs that also succeed from $H$. For each repair, Phase A caches a history after at most $10$ steps and shares it across tested methods and seeds. PRE and POST then continue from the same snapshot $D$, using the same agent, tools, environment, and eight-step limit; the repository's FAIL\_TO\_PASS tests judge both outcomes. An instance-method pair is \emph{preserved} when both continuations pass, \emph{damaged} when PRE passes and POST fails, a \emph{rescue} when PRE fails and POST passes, and \emph{unsolved} when both fail. The paired labels preserve the direction of each change: damage identifies a lost full-history success, while rescue identifies a success reached from the summary after full-history failure. Reporting these separately prevents improvements on some repairs from cancelling losses on others in an aggregate success rate. First-round damage and rescue cases receive two additional seeds. A backbone case counts as stable damage when POST fails in at least two of its three runs; the prompt comparison uses one run per case. Appendix~\ref{app:k3} reports the repeatability results.

\textbf{Recurring-Detail Retention} measures literal preservation of structured values that occur both before an artificial boundary in a never-compacted session and in the next $3$ to $8$ request payloads. Future payloads include retransmitted history, so recurrence denotes continued presence in the interaction. \textbf{Production-Chain Retention} scores details found in the next at most $5$ request payloads after each real compaction boundary, without requiring a match in the pre-boundary history. It also records whether a summary describes a completed file write as pending work. The two retention tracks use different boundaries for complementary purposes. Artificial boundaries allow the same recorded history to be summarized by multiple methods and scored against a common reference set. Production boundaries expose the summaries emitted during ongoing work, including successive replacements within one session. Task Preservation adds the execution outcome, so the benchmark reports both what remains in a summary and what the continuation accomplishes from it.

\subsection{Data and Construction}
\label{sec:data}

\begin{table*}[t]
\centering
\small
\renewcommand{\arraystretch}{1.15}
\begin{tabularx}{\textwidth}{@{}llX@{}}
\toprule
Evaluation & Size & Composition \\
\midrule
Task Preservation & 35 instances & Localized SWE-bench repairs from 7 repositories: pytest (18), sympy (6), xarray (3), pylint (3), requests (2), sphinx (2), and flask (1). \\
\midrule
Recurring-Detail Retention & 146 sessions & 143 real and 3 simulator-synthesized sessions; clients: Kimi Code (73), Claude Code (50), opencode (14), and other clients (9); 9{,}330 recurring details. \\
\midrule
Production-Chain Retention & 106 sessions & 224 production compaction boundaries; median 2 and maximum 4 boundaries per session. \\
\bottomrule
\end{tabularx}
\caption{Datasets for the three Compact-Bench evaluations.}
\label{tab:data}
\end{table*}

\paragraph{Task Preservation sample.}
The task set contains 35 localized SWE-bench repairs from $7$ repositories: pytest ($18$), sympy ($6$), xarray ($3$), pylint ($3$), requests ($2$), sphinx ($2$), and flask ($1$). Selection favors fixes confined to one test file, few FAIL\_TO\_PASS tests, and a median gold patch below $1$ KB. This concentrates evaluation on whether a continuation can turn the accumulated state into a specific repair. Each instance has a repository-specific environment recipe: Python 3.9 supports older repositories, while dependency constraints include \texttt{numpy<2} for xarray, era-matched \texttt{sphinxcontrib-*} packages for sphinx, and \texttt{py} for pylint. Inclusion requires the base repository to fail the FAIL\_TO\_PASS tests and the gold patch to pass them. Appendix~\ref{app:instances} provides instance-level tests and environment details.

\paragraph{Recurring-Detail Retention sample.}
This track uses $146$ sessions from coding-agent harness trajectory: $73$ Kimi Code, $50$ Claude Code, $14$ opencode, and $9$ other clients, including $3$ simulator-synthesized sessions. A candidate must have no detected compaction marker, at least $8$ requests, a maximum recorded request-token count of at least $60{,}000$, and a completeness score of at least $0.8$. Completeness is the fraction of eligible adjacent request pairs that pass the history-prefix and cache-continuity checks in Appendix~\ref{app:reprodetails}. If a session has $n$ requests, where $n$ is the request count, the artificial boundary is placed at zero-based request index $\lfloor 0.65n\rfloor$ (rounding down); at least $3$ requests must follow it, and at most $8$ are included in the future window. Client quotas and ordering are in the supplementary configuration record. We exclude sessions with fewer than three matching details across the pre-boundary history and future payloads. The final set has $9{,}330$ recurring details: paths, commit SHAs, pull-request identifiers, error types, URLs, key-value pairs, and IP addresses.

\paragraph{Production-Chain Retention sample.}
This track uses $106$ sessions from the same corpus, each with at least two confirmed production compactions. They contain $224$ boundaries: $106$ first, $106$ second, $10$ third, and $2$ fourth boundaries. At least one future detail can be evaluated at $140$ boundaries ($106$, $27$, $6$, and $1$ at the respective boundary positions); only these enter detail-retention scores.

\paragraph{Reference prompts.}
The prompt comparison uses $5$ production compaction prompts from coding-agent harnesses: Claude Code, Codex, Gemini CLI, opencode, and Kimi Code. They serve as reference methods in Task Preservation and Recurring-Detail Retention. Production-Chain Retention evaluates the summaries generated by the live systems in the recorded sessions.

\subsection{Scoring}
\label{sec:metrics}

\paragraph{Task Preservation.}
The damage rate is the fraction of PRE-pass cases that fail after compaction; the preservation rate is one minus this rate. Conditioning on full-history success identifies repairs that the continuation can complete before history replacement. The backbone comparison uses $22$ validated PRE-pass cases among $24$ evaluated cases; two are excluded because PRE failed in three re-run. We report preserved, damaged, rescued, and unsolved outcomes separately, using single-run results for the prompt comparison. For backbone cases selected for verification, stable damage requires POST failure in at least two of three runs.

\paragraph{Recurring-Detail Retention score.}
For any text $T$ (a history or request payload), let $A(T)$ be the deterministic extractor that returns distinct (type, value) pairs for paths, commit SHAs, pull-request or issue identifiers, error types, URLs, key-value pairs, and IP addresses. For session $i$, let $H_i$ be the pre-boundary history, $F_i$ the concatenation of the next three to eight request payloads (including retransmitted history), and $S_i$ the tested summary. The reference set $A_i=A(H_i)\cap A(F_i)$ contains details extracted from both texts. A pair matches if its literal value occurs as a case-sensitive substring in $S_i$. The per-session score is
\[
\operatorname{retention}_i = \frac{\text{matching pairs in }A_i}{|A_i|},
\]
where $|\cdot|$ denotes the number of elements in a set. Matching does not require a token boundary or the same type label in the summary. We report the unweighted mean of these session scores for each method; empty reference sets are excluded, and construction requires at least three reference details per session. Appendix~\ref{app:reprodetails} gives the extraction, normalization, and window rules.

\paragraph{Production-Chain Retention score.}
For session $i$ and its $k$th compaction boundary, let $F_{i,k}$ be the next at most five request payloads, $S_{i,k}$ the production summary, and $A_{i,k}=A(F_{i,k})$ the extracted future details. Here $i$ indexes sessions and $k$ is the boundary's position within a session. Unlike the recurring-detail reference set, $A_{i,k}$ does not require an intersection with pre-boundary history. As above, a detail pair matches when its value occurs as a case-sensitive substring in $S_{i,k}$. The score for a boundary is
\[
R_{i,k}=\frac{\text{matching detail pairs in }A_{i,k}}{|A_{i,k}|}.
\]
We average these scores separately for first, second, and later boundaries, using only boundaries with at least one extracted future detail. We also report the mean last-to-first ratio within sessions where both boundaries have scores and the first-boundary score is nonzero: 22 sessions meet the two-boundary criterion and 20 remain after excluding zero first-boundary scores. For state-time disorder, the detector flags a boundary if a basename matched by the successful-write patterns in pre-boundary history as created or modified appears in the production summary within 150 characters of a pending-work phrase. We report this flag rate by boundary position over all boundaries, including those without extractable future details. Appendix~\ref{app:reprodetails} specifies the matching patterns and window.

\subsection{Evaluation Validation}
\label{sec:quality}

We check three properties before scoring: \emph{solvability} (the gold patch passes in the supplied environment), \emph{discriminability} (the base repository fails the FAIL\_TO\_PASS tests), and \emph{protocol validity} (the agent and test judge use the same interpreter, and agent tests run against the live working tree).

\paragraph{Controlled replay.}
The agent and judge use the same per-instance interpreter, with PATH and PYTHONPATH pointing to the evaluation environment and live working tree. Phase A, PRE, and POST execute in the same directory, and the cached working-tree snapshot is restored before each continuation. This preserves the meaning of paths appearing in the history or summary and places both outcomes under the same conditions. Appendix~\ref{app:audit} reports validation experiments for these controls.

\paragraph{Deterministic endpoints.}
Task outcomes use repository tests, and retention scores use deterministic extraction and string matching. Reference values are identified from recorded histories and future requests before comparing summaries. These endpoints provide instance-level evidence: a changed test verdict identifies task damage, while a missing literal value identifies detail loss. Auxiliary human validation of decision and progress labels is documented separately in Appendix~\ref{app:audit}; these labels do not enter the reported retention scores.

\section{Results}
\label{sec:results}

\begin{table}[!tb]
\centering\small\setlength{\tabcolsep}{4pt}
\begin{tabular}{@{}lccc@{}}
\toprule
Summarizer & Round 1 & Stable & 95\% CI \\
\midrule
Claude Opus 4.8 & $5\%$ (1) & $5\%$ (1) & $1-22$ \\
Kimi-K3 & $18\%$ (4) & $18\%$ (4) & $7-39$ \\
GPT-5.6-Luna & $18\%$ (4) & $18\%$ (4) & $7-39$ \\
GPT-5.6-Sol & $27\%$ (6) & $27\%$ (6) & $13-48$ \\
Claude Fable 5 & $41\%$ (9) & $36\%$ (8) & $20-57$ \\
Claude Fable 5.1 & $36\%$ (8) & $36\%$ (8) & $20-57$ \\
GPT-6-Astra & $50\%$ (11) & $41\%$ (9) & $23-61$ \\
Qwen3.8-Max & $50\%$ (11) & $45\%$ (10) & $27-65$ \\
GPT-5.6-Terra & $45\%$ (10) & $45\%$ (10) & $27-65$ \\
Claude Opus 5.5 & $45\%$ (10) & $45\%$ (10) & $27-65$ \\
GLM-5.3 & $55\%$ (12) & $50\%$ (11) & $31-69$ \\
DeepSeek-V4-Pro & $55\%$ (12) & $55\%$ (12) & $35-73$ \\
Hy4-Preview & $64\%$ (14) & $59\%$ (13) & $39-77$ \\
DeepSeek-Flash & $64\%$ (14) & $64\%$ (14) & $43-80$ \\
\bottomrule
\end{tabular}
\caption{Task Preservation damage under the Claude Code prompt and Kimi-K3 continuation agent. Entries give the rounded percentage and count among $22$ validated PRE-pass cases. Stable damage requires POST failure in at least two of three runs; intervals are Wilson $95\%$ confidence intervals for that rate.}
\label{tab:backbone}
\end{table}

\subsection{Evaluation Setup}
\label{sec:setup}

The fourteen-backbone Task Preservation comparison fixes the Claude Code prompt and Kimi-K3 continuation agent; every method receives the same cached Phase-A history and repository snapshot. The prompt comparison fixes Kimi-K3 as summarizer and varies the five production prompts. Recurring-Detail Retention evaluates seven backbones with all five prompts, while Production-Chain Retention scores recorded summaries at live compaction boundaries. Damage rates use the $22$ validated PRE-pass repairs. In the backbone study, first-round damage or rescue receives two additional runs ($K{=}3$); both PRE and POST are rerun for Kimi-K3, while other backbones reuse PRE. Prompt comparisons use one run per case. Appendix~\ref{app:harness} specifies budgets, decoding settings, endpoints, and so on.

\subsection{Task Preservation Results}
\label{sec:task-preservation-results}

\paragraph{Backbone comparison.}
Table~\ref{tab:backbone} reports stable damage under the fixed Claude Code prompt. Claude Opus~$4.8$ preserves $21$ of the $22$ validated repairs, followed by Kimi-K3 and GPT-5.6-Luna with $18$ each. DeepSeek-Flash preserves $8$, yielding a $59$-percentage-point spread in damage rates across the fourteen summarizers ($5-64\%$). Because the history, repository state, and continuation agent are shared, this comparison measures the effect of the summarizer choice within a common repair setup. The table reports Wilson intervals alongside each rate.

Figure~\ref{fig:heatmap} reveals how the differences arise. The damage sets overlap partially: every backbone stably damages a repair that another preserves. Shared hard cases coexist with complementary successes. On pytest-10081, Claude Opus~$4.8$ preserves the repair while the other thirteen backbones damage it; on sphinx-9230, nine backbones, including Claude Opus~$4.8$, damage the repair and five preserve it. Instance-level outcomes therefore identify both widely shared failure cases and successful alternatives for the same task.

\begin{figure}[!tb]
\centering
\input{cell_heatmap_tikz}
\caption{POST outcomes on the 22 validated PRE-pass cells ($r_1$) for the fourteen summarizer backbones. P: POST passes; F: POST fails. Full model names are listed in Table~\ref{tab:backbone}. Damage sets overlap only partially.}
\label{fig:heatmap}
\end{figure}

\paragraph{Agreement across repairs.}
The fourteen backbones agree on four of the $22$ validated repairs and differ on the remaining $18$ (Appendix~\ref{app:grid}). Every validated pytest repair and two of the three xarray repairs are damaged by at least one backbone, while the validated requests and pylint cases remain preserved throughout. Claude Opus~$4.8$ preserves $17$ of the $18$ repairs with disagreement. Its advantage is thus distributed across the disagreement set, with sphinx-9230 identifying a concrete case for which other summarizers supply a successful continuation state. The paired grid is useful precisely because these alternatives start from the same history and snapshot. A difficult repair need not fail under every summary: pytest-10081 has one preserving backbone, and sphinx-9230 has five. The benchmark thus localizes the configuration choices associated with success, while the aggregate rates quantify how frequently each choice preserves the tested repairs.

\paragraph{Effect of repeated runs.}
Repeated runs confirm $120$ of the $126$ first-round damaged instance--backbone pairs. GPT-6-Astra changes from $11$ first-round damaged repairs to $9$ confirmed cases; Claude Fable~$5$, Qwen3.8-Max, GLM-$5.3$, and Hy4-Preview each change by one. Claude Opus~$4.8$ and DeepSeek-Flash remain the lowest- and highest-damage backbones. The broad spread between summarizers therefore persists under the confirmation rule, and most observed damage recurs across independent samples. Appendix~\ref{app:k3} reports all observed outcomes for the selected cases.

\paragraph{Prompt comparison.}
With Kimi-K3 as summarizer, Kimi Code produces the most POST passes ($19/22$), followed by Claude Code ($18/22$), Gemini CLI ($17/22$), Codex ($16/22$), and opencode ($16/22$) (Table~\ref{tab:variants}). On the separate 146-session retention set, Claude Code retains the largest share of recurring details ($15.0\%$). The two evaluations distinguish a prompt that best supports completing the sampled repairs from one that best preserves recurring details in the session corpus. This difference makes the evaluation target consequential when selecting a compaction prompt.

\begin{table}[!tb]
\centering
\small
\setlength{\tabcolsep}{5pt}
\begin{tabular}{@{}lcc@{}}
\toprule
Prompt & POST pass $\uparrow$ & Detail retention $\uparrow$ \\
\midrule
Kimi Code & 19/22 = 86\% & 8.2\% \\
Claude Code & 18/22 = 82\% & 15.0\% \\
Gemini CLI & 17/22 = 77\% & 7.2\% \\
Codex & 16/22 = 73\% & 9.5\% \\
opencode & 16/22 = 73\% & 9.5\% \\
\bottomrule
\end{tabular}
\caption{Task Preservation POST-pass counts among 22 validated PRE-pass instances and Recurring-Detail Retention mean scores over the separate 146-session set, both using Kimi-K3 as summarizer.}
\label{tab:variants}
\end{table}

The prompt grid separates shared failures from prompt-specific losses (Appendix~\ref{app:variantgrid}). Sphinx-9230 fails under all five prompts, and pytest-7571 fails under four. In contrast, xarray-3677, pylint-6903, and pytest-7521 fail under opencode alone; pytest-7490 fails under Claude Code alone. Kimi Code and Claude Code also have different failure profiles: Kimi Code preserves the three pytest repairs damaged by Claude Code, while Claude Code preserves requests-5414 and sympy-13551, which Kimi Code damages. Sphinx-9230 is their shared failure. These profiles explain how the two leading prompts differ beyond their one-repair aggregate gap. The controlled prompt comparison also holds the summarizer fixed, demonstrating that the compaction instruction alone is a consequential configuration choice in this setup. Its case-level outcomes complement the backbone experiment, which holds the instruction fixed. Together, the two comparisons separate model choice from prompt choice while retaining the same continuation and executable criterion.

\subsection{Detail Retention Results}
\label{sec:detail-retention-results}

\paragraph{Recurring-Detail Retention.}
On 146 sessions, DeepSeek-Flash attains the highest LLM retention score ($16.0\%$), followed by Kimi-K3 ($15.0\%$) under the Claude Code prompt (Table~\ref{tab:detail-retention}). Across the seven-backbone, five-prompt matrix, scores range from $5.0\%$ to $16.0\%$. DeepSeek-Flash leads under Claude Code, Gemini CLI, and Kimi Code, while GPT-5.6-Terra leads under Codex and opencode. The matrix therefore reveals an interaction between summarizer and prompt: the same backbone need not lead under every compaction instruction. Appendix~\ref{app:detail-retention-client} supplies the complete matrix and matched client-subset comparisons.

\begin{table}[!tb]
\centering
\footnotesize
\begin{tabular}{@{}lc@{}}
\toprule
Backbone & Detail-retention score $\uparrow$ \\
\midrule
DeepSeek-Flash & 16.0\% \\
Kimi-K3 & 15.0\% \\
GPT-5.6-Sol & 14.8\% \\
GPT-5.6-Luna & 13.3\% \\
DeepSeek-V4-Pro & 11.9\% \\
GLM-5.3 & 9.5\% \\
GPT-5.6-Terra & 8.2\% \\
\midrule
\texttt{anchor\_index} (mechanical) & 79.1\% \\
\bottomrule
\end{tabular}
\caption{Recurring-Detail Retention score by summarizer backbone (Claude Code prompt). Per-backbone valid $n$ is 144--146 of 146 sessions. API or parse errors are excluded, while empty or whitespace summaries score 0 and remain included.}
\label{tab:detail-retention}
\end{table}

\paragraph{What the retention gap measures.}
The mechanical \texttt{anchor\_index} reference retains $79.1\%$ of recurring details, a $63.1$-percentage-point gap over the best LLM summary. These values are directly recoverable from recorded history under a bounded extraction procedure: the reference reads at most $400{,}000$ serialized characters and keeps at most $40$ values per type. The comparison quantifies how much recurring structured information survives extraction versus generated summarization. Task Preservation supplies the complementary execution test, assessing whether the resulting continuation state supports completing a repair. The retention reference gives each recurring value an explicit membership test, making omissions inspectable at the level of paths, identifiers, error types, or other extracted strings. Scores are averaged per session, so a session with many labels does not dominate the corpus mean. This consistently compares histories with different amounts of structured information.

The joint results distinguish detail retention from task preservation. DeepSeek-Flash has the highest retention score under Claude Code but damages $14$ of $22$ repairs, compared with $4$ for Kimi-K3. With Kimi-K3 fixed, Claude Code retains $15.0\%$ of recurring details versus $8.2\%$ for Kimi Code, while Kimi Code preserves one more repair. These are configuration-level comparisons across separate datasets. They show why evaluating both properties is useful: a configuration that preserves more recurring strings can still supply a less effective state for task continuation.

The prompts also impose different output structures. Claude Code requests nine sections and produces summaries of about $10{,}000$ characters on the retention set, compared with $1{,}600-5{,}200$ for the other prompts. Kimi Code requests a first-person handoff focused on continuing the task. Thus, the observed prompt profiles combine differences in instructions, organization, and output length. Compact-Bench evaluates these production configurations as a whole, exposing how their information-retention and task-preservation outcomes differ.

\begin{figure}[!tb]
\centering
\includegraphics[trim=2bp 2bp 2bp 2bp,clip,width=7.6cm,height=4.55cm]{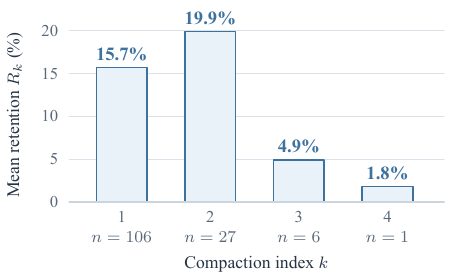}
\caption{Mean retention $R_k$ per compaction index across 106 production compaction chains. Sample sizes shrink quickly with $k$.}
\label{fig:chain-retention}
\end{figure}

\paragraph{Production-Chain Retention.}
Across 106 live sessions, mean retention is $15.7\%$ at the first boundary ($n{=}106$) and $19.9\%$ at the second ($n{=}27$); third- and fourth-boundary means are $4.9\%$ ($n{=}6$) and $1.8\%$ ($n{=}1$) (Figure~\ref{fig:chain-retention}). Each mean summarizes the evaluable subset at that position. The production measurements extend the benchmark from an artificial boundary to the summaries that agents actually receive during continued work. They document recurring-detail coverage across repeated history replacements, with per-session ratios and instruction-conditioned breakdowns in Appendix~\ref{app:detail-retention-client}. The first-boundary result covers all $106$ sessions and places the controlled retention measurements alongside a deployed-use reference. Subsequent boundaries retain their own future-request windows, capturing the details present in the interaction after each replacement. This boundary-specific view preserves the context in which each production summary is used, rather than merging successive summaries into a single session-level artifact.

\subsection{Diagnosing Task Damage}
\label{sec:analysis}

Table~\ref{tab:retained-fix-evidence} examines four stable Kimi-K3 failures. Every summary retains its target file; the pytest summaries also contain $8/12$, $10/12$, and $5/5$ audited gold-patch tokens. Pytest-7571 is particularly informative: all five audited tokens survive, yet the continuation fails. Sphinx-9230 similarly retains strings from the fix but omits the reasoning that \texttt{split(None, 1)} must become \texttt{rsplit}. The cases expose a distinction between preserving repair ingredients and preserving the progress that connects them to a completed edit.

\begin{table}[!tb]
\centering
\footnotesize
\begin{tabular}{@{}lcc@{}}
\toprule
Damaged cell & Target file in $S$ & Fix evidence in $S$ \\
\midrule
pytest-10081 & yes (1/1) & 8/12 \\
pytest-7490 & yes (1/1) & 10/12 \\
pytest-7571 & yes (1/1) & 5/5 \\
sphinx-9230 & yes (1/1) & 4 listed strings \\
\bottomrule
\end{tabular}
\caption{Retained repair evidence in the four stably damaged cells under Kimi-K3 summaries. The target file appears in all four; fractions give retained gold-patch tokens for the pytest cases. The sphinx row lists four strings from the worked example and has no exhaustive token denominator.}
\label{tab:retained-fix-evidence}
\end{table}

\paragraph{From evidence to action.}

The sphinx-9230 trace makes that distinction visible in subsequent actions. PRE reads the target region, builds and checks a minimal reproduction, locates the call, and applies a short Python edit. POST reaches the relevant file, then searches broadly and reads test and reproduction code without making the repair. The trace contains $8$ PRE tool actions and $13$ POST actions within the shared eight-step budget, since a step can contain multiple tool actions (Appendix~\ref{app:trace}). Counts of explicit \texttt{read\_file} calls show at most one repeat read per condition in each damaged case (Appendix~\ref{app:redo}). The observed divergence is in the transition from relevant evidence to an effective edit, even with the target file still accessible.

This diagnosis illustrates the role of paired continuation in the benchmark. Retention scores locate missing details; paired traces reveal how a preserved or altered working state unfolds into subsequent actions. On the inspected repairs, the decisive outcome is executable task completion, and paths or code tokens provide evidence for interpreting that outcome. Combining these views makes it possible to distinguish a missing identifier from a continuation that still has the identifier but does not complete the repair.

\paragraph{Evaluation cost.}
The complete 35-repair study required about 8{,}000 model calls, with compression accounting for under $5\%$ of tokens. The main computational cost lies in generating and judging continuations, which provide the executable evidence for task preservation.

Appendix~\ref{app:harness} provides scoring rules, configurations, environment recipes, and instance-level outcome tables, supporting inspection of how each aggregate follows from its underlying cases.

\section{Conclusion}
\label{sec:conclusion}

Compact-Bench measures what survives context compaction through paired task continuation, recurring-detail matching, and real production chains. Its shared-boundary comparison separates summarizers by their ability to preserve executable repair progress: Claude Opus~$4.8$ has the lowest stable damage, with a $5-64\%$ range across fourteen backbones. Prompt choice also matters, with Kimi Code preserving the most repairs and Claude Code retaining the most recurring details under Kimi-K3. The paired failures establish the benchmark's central distinction: retaining paths and code strings can coexist with losing the progress needed to finish a task. Evaluating both information and subsequent action makes that loss visible.

\section*{Limitations}
\label{sec:limitations}

Results are conditional on the tested repository repairs, continuation agent, and budget. Detail-retention scores capture literal recurrence, while later production-boundary estimates use small, changing session subsets. Private session payloads and cached repair states are unavailable; Appendix~\ref{app:harness} details the released scoring rules, configurations, and outcome tables.

\section*{Ethical Considerations}
\label{sec:ethics}

The Recurring-Detail Retention and Production-Chain Retention evaluations use real agentic sessions collected from online agent traffic on the authors' institutional platform under its existing data-use agreements; sessions were analyzed in place and were not solicited from users for this study. The data was processed to extract structural items (paths, identifiers, error types) for labeling. The paper reports aggregate retention statistics; private session payloads and session-level label values are not distributed. Paths, URLs, and IP addresses can identify users or infrastructure; the distributed artifacts therefore exclude these private session-level values. Task Preservation reuses public SWE-bench instances under their original licenses. All reported experiments were run through metered API gateways under the authors' institutional accounts. The benchmark is intended to help identify and reduce compaction-induced failures in deployed coding agents.

\bibliography{custom}

\appendix
\renewcommand{\floatpagefraction}{0.65}
\renewcommand{\dblfloatpagefraction}{0.65}

\section{First-Round Task Preservation Outcomes}
\label{app:grid}
Table~\ref{tab:full_grid} lists first-round outcomes for all fourteen backbones. X marks pytest-10051 and pytest-7236, excluded because PRE fails in three independent environments. Dashes mark cells outside the cross-backbone PRE-pass comparison. The caption defines the outcome codes and model abbreviations.

\input{grid_table}

\section{Three-Run Task Preservation Verification}
\label{app:k3}
Table~\ref{tab:k3} gives the observed runs for all 126 first-round damaged instance--backbone pairs: 120 receive the stable-damage label SD and six the variability label V. The caption specifies which conditions are resampled.

\input{k3_table}

\section{Detail Retention Breakdowns}
\label{app:detail-retention-client}

\paragraph{Backbone, prompt, and client.}
Table~\ref{tab:detail-retention-matrix} reports all seven-backbone, five-prompt scores; Tables~\ref{tab:detail-retention-client-k3} and~\ref{tab:detail-retention-client-gpt} break down Kimi-K3 and GPT-5.6-Luna results by client source. Under Kimi-K3 with Claude Code, Kimi Code sessions average $21.4\%$, Claude Code $9.4\%$, and opencode $6.6\%$. The client groups have different task mixes and detail densities; comparisons within each group hold that sampled subset fixed. Claude Code ranks first within most groups, consistent with its pooled score. The labels \texttt{fake-kimi} and \texttt{fake-cc} identify two and one simulator-generated sessions, respectively; \texttt{kilocode} and \texttt{hertz} are smaller client groups, and \texttt{other} groups the remaining sources. The zero scores for \texttt{fake-cc} and \texttt{hertz} each correspond to one session with no retained measured detail. Valid sample counts range from 144 to 146 after API or parse failures are excluded.

\paragraph{Production boundaries.}
Of 224 production boundaries, 140 contain evaluable future details. Among these, the 84 with an explicit summary instruction average $19.0\%$ retention, and the 56 without one average $11.4\%$. This groups recorded boundaries by instruction presence; the underlying sessions and compaction conditions also vary. Twenty-two sessions have scores at both their first and last boundaries; excluding two zero first-boundary scores leaves 20 sessions with a mean last-to-first ratio of $4.07$. The ratio describes this conditional subset. The completed-write detector flags zero boundaries under the matching rules in Appendix~\ref{app:reprodetails}.

\input{detail_retention_matrix}

\section{Evaluation Validation Results}
\label{app:audit}

\paragraph{Interpreter and directory controls.}
Table~\ref{tab:audit} compares three harness settings: the system interpreter (v1), the instance interpreter with separate execution directories (v2), and the instance interpreter with a shared directory and snapshot restoration (v3). The benchmark uses v3. Interpreter alignment makes older repositories importable to both agent and judge; for example, sympy-13480 changes from unsolved to preserved. Directory alignment preserves the meaning of paths copied into summaries. In the separate-directory pilot, copied Phase-A paths redirected POST commands outside the judged tree in 22 of 34 cases. Sharing and restoring the directory resolves this mismatch: xarray-6721 and pytest-7521 change from damage to preserved, while sphinx-9230 remains damaged. These comparisons validate the environment and state controls used in the paired evaluation.

\paragraph{Auxiliary label validation.}
Decision and progress labels were examined separately from the deterministic retention metric. From 1{,}747 raw candidates, many originating in system prompts and skill documents, a stratified sample of 60 items (20 decisions and 40 progress items) was independently reviewed by a human and Kimi-K3. They agreed on 59 items ($98.3\%$). One disagreement concerned a literal duplicate and was adjudicated in favor of the LLM annotation. The resulting auxiliary gold set contains 31 items from eight sessions. This review validates the auxiliary labels; the headline retention scores use the structured-value extraction rules in Appendix~\ref{app:reprodetails}.

\input{audit_table}

\section{Task Preservation Instances}
\label{app:instances}

Table~\ref{tab:instances} records the 35 repairs, target-test counts (median 1, maximum 10), and environments used for base-fail and gold-pass validation. Older repositories use Python 3.9 when they cannot import under Python 3.10 or later. Filtering excluded four django candidates because of environment cost and one pytest candidate whose assertion depended on pytest internals.

\input{instances_table}

\section{Task Preservation Outcomes by Prompt}
\label{app:variantgrid}

Table~\ref{tab:variant_grid} reports first-round outcomes with Kimi-K3 and five production prompts: 22 validated PRE-pass repairs and two excluded cases marked X. The grid separates shared failures from prompt-specific losses; Kimi Code's three failures are requests-5414, sphinx-9230, and sympy-13551.

\input{variant_grid_table}

\section{Harness and Reproducibility}
\label{app:harness}

\paragraph{Artifact licenses.}
The reference implementation is distributed under the project's Apache License~2.0. Upstream repository licenses remain separate: pytest uses MIT; requests and xarray use Apache~2.0; Flask and SymPy use BSD~3-Clause; Sphinx uses BSD~2-Clause; and pylint distributes GPL~2.0 license terms. Incorporated components may carry additional notices. The supplement's license manifest links the upstream license files at the instance commits. These licenses do not grant access to private sessions or relicense upstream issue text.

\paragraph{Availability and reproducibility.}
The accompanying \texttt{compact-bench-reproducibility.zip} supplies reference extraction and scoring code, transcript preprocessing, coding-agent inputs and tool schemas, environment recipes, and file schemas described in Appendix~\ref{app:reprodetails}. The Task Preservation first-round grids and selected additional runs (Appendices~\ref{app:grid}, \ref{app:k3}, and \ref{app:variantgrid}) support recomputation of the reported task rates. The package excludes private session payloads, historical candidate order and sampling budget, and cached Task Preservation histories and working-tree snapshots. Recomputing session-level aggregates requires the source payloads; fresh Task Preservation runs generate new trajectories. The released rules and configurations specify the operations for implementing those evaluations.

\paragraph{Agent and tools.} The continuation agent runs with two tools: \texttt{read\_file} (confined to the working directory) and \texttt{run\_command} with an allowlist (\texttt{pytest}, \texttt{python}, \texttt{grep}, \texttt{rg}, \texttt{cat}, \texttt{sed}, \texttt{head}, \texttt{tail}, \texttt{find}, \texttt{echo}, \texttt{cd}, \texttt{pwd}, \texttt{wc}, \texttt{diff}, \texttt{patch}, \texttt{git}, \texttt{mkdir}, \texttt{cp}, \texttt{mv}, \texttt{touch}, \texttt{pip}, \texttt{rm}, \texttt{ls}). Command outputs are truncated to 3{,}000 characters; file reads to 8{,}000. The agent's shell receives the instance venv on PATH and the working directory on PYTHONPATH, so its \texttt{python} and \texttt{pytest} invocations exercise the live working tree with the judging interpreter.

\paragraph{Budgets and caching.} Phase A allows at most 10 steps; each continuation at most 8; command timeout is 120s and test timeout is 300s. Compression inputs are preprocessed before the overall cap: in Task Preservation and cross-backbone runs, each text block is truncated to 2{,}000 characters and each tool input or output to 200 characters; in Recurring-Detail Retention, the limits are 4{,}000 and 500 characters; the serialized transcript is then capped at its last 250{,}000 characters. The \texttt{anchor\_index} reference extracts from the first 400{,}000 characters of serialized history and keeps at most 40 values per type after sorting. Compression outputs are capped at 6{,}144 tokens for Kimi-K3 main runs and 16{,}384 tokens for cross-backbone and Recurring-Detail Retention runs. Each phase-A trajectory is cached and shared across methods and seeds; the working-directory snapshot is restored between PRE and POST.

\paragraph{Models and endpoints.} Kimi-K3 and Claude-* models are called through the Anthropic messages endpoint; GPT-* models through the OpenAI chat-completions or responses endpoint, all via the same metered gateway with per-family API keys. Runs reported here were executed in September--October 2026 with gateway model identifiers \texttt{kimi-k3}~\citep{kimi2026k3}, \texttt{claude-opus-4-8}~\citep{anthropic2026opus48}, \texttt{claude-fable-5}~\citep{anthropic2026fable5}, \texttt{claude-fable-5-1}~\citep{anthropic2026fable51}, \texttt{gpt-5.6-luna}~\citep{openai2026gpt56}, \texttt{gpt-5.6-sol}~\citep{openai2026gpt56}, \texttt{gpt-5.6-terra}~\citep{openai2026gpt56}, \texttt{gpt-6-astra}~\citep{openai2026gpt6astra}, \texttt{claude-opus-5-5}~\citep{anthropic2026opus55}, \texttt{qwen3.8-max}~\citep{qwen2026qwen38max}, \texttt{glm-5.3}~\citep{zai2026glm53}, \texttt{deepseek-v4-pro}~\citep{deepseek2026v4,deepseek2026v4proga}, \texttt{hy4-preview}~\citep{tencent2026hy4preview}, and \texttt{deepseek-flash}~\citep{deepseek2026v41flash,deepseek2026flashrelease}. Kimi-K3 runs with thinking disabled; thinking-style models run with default settings. Continuation turns use \texttt{max\_tokens} 4{,}096; compression output caps follow the preceding paragraph (6{,}144 for Kimi-K3 main runs, 16{,}384 for cross-backbone and Recurring-Detail Retention runs). The gateway offers no seed parameter, so seeds are independent server-side samples; temperature and other decoding settings are the service defaults, identical across conditions. Per-family environment pins follow fixed recipes: \texttt{numpy<2}, \texttt{pandas<2}, \texttt{setuptools\_scm<8} for xarray; \texttt{py} for pylint; \texttt{docutils<0.18} and era-pinned \texttt{sphinxcontrib-applehelp==1.0.2}, \texttt{-devhelp==1.0.2}, \texttt{-htmlhelp==2.0.0}, \texttt{-qthelp==1.0.3}, \texttt{-serializinghtml==1.1.5} for sphinx; plus \texttt{pytest<8}, \texttt{werkzeug<3}, and \texttt{urllib3<2} everywhere. The pytest source-archive version shim sets \texttt{SETUPTOOLS\_\allowbreak SCM\_\allowbreak PRETEND\_\allowbreak VERSION\_\allowbreak FOR\_\allowbreak PYTEST} to \texttt{7.99.0} during installation and execution. The supplement records installation order and dependency version constraints.

\paragraph{AI assistance.}
Generative AI tools assisted with literature search, manuscript restructuring and language editing, and LaTeX and figure preparation. The authors remain responsible for verifying the cited sources, factual statements, and interpretations of experimental results.

\paragraph{Judging.} FAIL\_TO\_PASS tests are normalized to pytest node IDs; for repositories whose FAIL\_TO\_PASS entries are bare test function names (sympy), the target file is located by searching the files touched by the test patch for the function definition. Judging reinstalls the working tree non-editable into the venv before running tests, so both source-layout and flat-layout repositories are scored against the live tree.

\section{Production Compaction Prompts}
\label{app:prompts}

The five production prompts are included verbatim below. The Claude Code prompt (634 characters) reads in full:

\begin{quote}\small\ttfamily\obeylines
Your task is to create a detailed summary of the conversation so far, paying close attention to the user's explicit requests and your previous actions.
This summary should be thorough in capturing technical details, code patterns, and architectural decisions that would be essential for continuing development work without losing context.
Your summary should include the following sections:
1. Primary Request and Intent
2. Key Technical Concepts
3. Files and Code Sections (include full code snippets where applicable)
4. Errors and fixes
5. Problem Solving
6. All user messages
7. Pending Tasks
8. Current Work
9. Optional Next Step
\end{quote}

The opencode prompt (537 characters) reads in full:

\begin{quote}\small\ttfamily\obeylines
You are an anchored context summarization assistant for coding sessions.
Summarize only the conversation history you are given. The newest turns may be kept verbatim outside your summary, so focus on the older context that still matters for continuing the work.
Always follow the exact output structure requested by the user prompt. Keep every section, preserve exact file paths and identifiers when known, and prefer terse bullets over paragraphs.
Do not answer the conversation itself. Respond in the same language as the conversation.
\end{quote}

The Codex prompt (423 characters) reads in full:

\begin{quote}\small\ttfamily\obeylines
You are performing a CONTEXT CHECKPOINT COMPACTION. Create a handoff summary for another LLM that will resume the task.
Include:
- Current progress and key decisions made
- Important context, constraints, or user preferences
- What remains to be done (clear next steps)
- Any critical data, examples, or references needed to continue
Be concise, structured, and focused on helping the next LLM seamlessly continue the work.
\end{quote}

The Kimi Code prompt (296 characters) reads in full:

\begin{quote}\small\ttfamily\obeylines
You are about to run out of context. Write a first-person handoff note to yourself so you can seamlessly continue this task after the earlier conversation is cleared.
--- This message is a direct task, not part of the above conversation ---
Write the note as your own continuing train of thought.
\end{quote}

The Gemini CLI prompt (279 characters) reads in full:

\begin{quote}\small\ttfamily\obeylines
You are an expert context compressor. Your job is to drastically shorten the following conversation while preserving the absolute core semantic meaning, facts, and intent. Omit all conversational filler, pleasantries, or redundant information. Return ONLY the compressed summary.
\end{quote}

\section{Worked Example: Action Traces}
\label{app:trace}

Table~\ref{tab:trace} presents the paired sphinx-9230 continuation discussed in Section~\ref{sec:analysis}. The trace follows the transition from inspecting the target region to constructing a repair. PRE locates and edits the relevant call; POST accesses the same file and continues exploring tests and reproduction code until its continuation budget is exhausted.

\input{trace_table}

\section{Repeated File Reads}
\label{app:redo}

\input{redo_table}

A repeated read is an explicit \texttt{read\_file} call targeting a file already read in Phase A; \texttt{sed}, \texttt{cat}, and Python command reads are outside this count. Table~\ref{tab:redo} records at most one per condition in each damaged case. Table~\ref{tab:trace} supplies the complementary action-level view for sphinx-9230.

\section{Data Schemas}
\label{app:schema}

\paragraph{Task Preservation instance record.} \texttt{instance\_id}, \texttt{repo}, \texttt{base\_commit}, \texttt{problem\_statement}, \texttt{FAIL\_TO\_PASS}, \texttt{PASS\_TO\_PASS}, \texttt{test\_patch}, \texttt{patch} (gold patch), \texttt{version}, \texttt{environment\_setup\_commit}.

\paragraph{Recurring-Detail Retention sample record.} In \texttt{samples.jsonl}: \texttt{sample\_id}, \texttt{sess}, \texttt{client}, \texttt{boundary\_idx} (zero-based request index), \texttt{n\_reqs}, \texttt{H} (boundary-request messages), \texttt{F\_len}, \texttt{atoms} (the schema field for extracted recurring-detail labels, stored as \texttt{\{kind, value\}} pairs), \texttt{n\_atoms}, and \texttt{future\_req\_ids}. The \texttt{future\_req\_ids} field references full payloads in the private source data; these payloads are stored separately from the sample record.

\paragraph{Production-Chain Retention result record.} The stored result record contains \texttt{sess}, \texttt{client}, \texttt{n\_bounds}, \texttt{per\_k}, and \texttt{ratio\_last\_first}. Each \texttt{per\_k} entry contains \texttt{k}, \texttt{marker}, \texttt{form}, \texttt{n\_atoms\_F}, \texttt{R\_k}, \texttt{triplet\_errors}, and \texttt{S\_len}. Summary text, extracted detail values, and successful-write matches are intermediate values and are not saved in this result schema. The supplement provides machine-readable schemas for these files and the Recurring-Detail Retention scoring outputs.

\input{reproducibility_details}

\end{document}

%% file: cell_heatmap_tikz.tex
\begin{tikzpicture}[x=10.8pt,y=10.8pt,font=\fontsize{8}{9.6}\selectfont]
\node[anchor=east,inner sep=2pt] at (0,-0.5) {flask-5014};
\filldraw[fill=teal!16,draw=white,line width=.35pt] (0,0) rectangle (1,-1);
\node[inner sep=0pt] at (0.5,-0.5) {P};
\filldraw[fill=teal!16,draw=white,line width=.35pt] (1,0) rectangle (2,-1);
\node[inner sep=0pt] at (1.5,-0.5) {P};
\filldraw[fill=teal!16,draw=white,line width=.35pt] (2,0) rectangle (3,-1);
\node[inner sep=0pt] at (2.5,-0.5) {P};
\filldraw[fill=teal!16,draw=white,line width=.35pt] (3,0) rectangle (4,-1);
\node[inner sep=0pt] at (3.5,-0.5) {P};
\filldraw[fill=orange!28,draw=white,line width=.35pt] (4,0) rectangle (5,-1);
\node[inner sep=0pt] at (4.5,-0.5) {F};
\filldraw[fill=teal!16,draw=white,line width=.35pt] (5,0) rectangle (6,-1);
\node[inner sep=0pt] at (5.5,-0.5) {P};
\filldraw[fill=orange!28,draw=white,line width=.35pt] (6,0) rectangle (7,-1);
\node[inner sep=0pt] at (6.5,-0.5) {F};
\filldraw[fill=teal!16,draw=white,line width=.35pt] (7,0) rectangle (8,-1);
\node[inner sep=0pt] at (7.5,-0.5) {P};
\filldraw[fill=orange!28,draw=white,line width=.35pt] (8,0) rectangle (9,-1);
\node[inner sep=0pt] at (8.5,-0.5) {F};
\filldraw[fill=teal!16,draw=white,line width=.35pt] (9,0) rectangle (10,-1);
\node[inner sep=0pt] at (9.5,-0.5) {P};
\filldraw[fill=teal!16,draw=white,line width=.35pt] (10,0) rectangle (11,-1);
\node[inner sep=0pt] at (10.5,-0.5) {P};
\filldraw[fill=orange!28,draw=white,line width=.35pt] (11,0) rectangle (12,-1);
\node[inner sep=0pt] at (11.5,-0.5) {F};
\filldraw[fill=orange!28,draw=white,line width=.35pt] (12,0) rectangle (13,-1);
\node[inner sep=0pt] at (12.5,-0.5) {F};
\filldraw[fill=orange!28,draw=white,line width=.35pt] (13,0) rectangle (14,-1);
\node[inner sep=0pt] at (13.5,-0.5) {F};
\node[anchor=east,inner sep=2pt] at (0,-1.5) {requests-5414};
\filldraw[fill=teal!16,draw=white,line width=.35pt] (0,-1) rectangle (1,-2);
\node[inner sep=0pt] at (0.5,-1.5) {P};
\filldraw[fill=teal!16,draw=white,line width=.35pt] (1,-1) rectangle (2,-2);
\node[inner sep=0pt] at (1.5,-1.5) {P};
\filldraw[fill=teal!16,draw=white,line width=.35pt] (2,-1) rectangle (3,-2);
\node[inner sep=0pt] at (2.5,-1.5) {P};
\filldraw[fill=teal!16,draw=white,line width=.35pt] (3,-1) rectangle (4,-2);
\node[inner sep=0pt] at (3.5,-1.5) {P};
\filldraw[fill=teal!16,draw=white,line width=.35pt] (4,-1) rectangle (5,-2);
\node[inner sep=0pt] at (4.5,-1.5) {P};
\filldraw[fill=teal!16,draw=white,line width=.35pt] (5,-1) rectangle (6,-2);
\node[inner sep=0pt] at (5.5,-1.5) {P};
\filldraw[fill=teal!16,draw=white,line width=.35pt] (6,-1) rectangle (7,-2);
\node[inner sep=0pt] at (6.5,-1.5) {P};
\filldraw[fill=teal!16,draw=white,line width=.35pt] (7,-1) rectangle (8,-2);
\node[inner sep=0pt] at (7.5,-1.5) {P};
\filldraw[fill=teal!16,draw=white,line width=.35pt] (8,-1) rectangle (9,-2);
\node[inner sep=0pt] at (8.5,-1.5) {P};
\filldraw[fill=teal!16,draw=white,line width=.35pt] (9,-1) rectangle (10,-2);
\node[inner sep=0pt] at (9.5,-1.5) {P};
\filldraw[fill=teal!16,draw=white,line width=.35pt] (10,-1) rectangle (11,-2);
\node[inner sep=0pt] at (10.5,-1.5) {P};
\filldraw[fill=teal!16,draw=white,line width=.35pt] (11,-1) rectangle (12,-2);
\node[inner sep=0pt] at (11.5,-1.5) {P};
\filldraw[fill=teal!16,draw=white,line width=.35pt] (12,-1) rectangle (13,-2);
\node[inner sep=0pt] at (12.5,-1.5) {P};
\filldraw[fill=teal!16,draw=white,line width=.35pt] (13,-1) rectangle (14,-2);
\node[inner sep=0pt] at (13.5,-1.5) {P};
\node[anchor=east,inner sep=2pt] at (0,-2.5) {xarray-3677};
\filldraw[fill=teal!16,draw=white,line width=.35pt] (0,-2) rectangle (1,-3);
\node[inner sep=0pt] at (0.5,-2.5) {P};
\filldraw[fill=teal!16,draw=white,line width=.35pt] (1,-2) rectangle (2,-3);
\node[inner sep=0pt] at (1.5,-2.5) {P};
\filldraw[fill=teal!16,draw=white,line width=.35pt] (2,-2) rectangle (3,-3);
\node[inner sep=0pt] at (2.5,-2.5) {P};
\filldraw[fill=orange!28,draw=white,line width=.35pt] (3,-2) rectangle (4,-3);
\node[inner sep=0pt] at (3.5,-2.5) {F};
\filldraw[fill=orange!28,draw=white,line width=.35pt] (4,-2) rectangle (5,-3);
\node[inner sep=0pt] at (4.5,-2.5) {F};
\filldraw[fill=teal!16,draw=white,line width=.35pt] (5,-2) rectangle (6,-3);
\node[inner sep=0pt] at (5.5,-2.5) {P};
\filldraw[fill=teal!16,draw=white,line width=.35pt] (6,-2) rectangle (7,-3);
\node[inner sep=0pt] at (6.5,-2.5) {P};
\filldraw[fill=teal!16,draw=white,line width=.35pt] (7,-2) rectangle (8,-3);
\node[inner sep=0pt] at (7.5,-2.5) {P};
\filldraw[fill=teal!16,draw=white,line width=.35pt] (8,-2) rectangle (9,-3);
\node[inner sep=0pt] at (8.5,-2.5) {P};
\filldraw[fill=teal!16,draw=white,line width=.35pt] (9,-2) rectangle (10,-3);
\node[inner sep=0pt] at (9.5,-2.5) {P};
\filldraw[fill=teal!16,draw=white,line width=.35pt] (10,-2) rectangle (11,-3);
\node[inner sep=0pt] at (10.5,-2.5) {P};
\filldraw[fill=teal!16,draw=white,line width=.35pt] (11,-2) rectangle (12,-3);
\node[inner sep=0pt] at (11.5,-2.5) {P};
\filldraw[fill=orange!28,draw=white,line width=.35pt] (12,-2) rectangle (13,-3);
\node[inner sep=0pt] at (12.5,-2.5) {F};
\filldraw[fill=orange!28,draw=white,line width=.35pt] (13,-2) rectangle (14,-3);
\node[inner sep=0pt] at (13.5,-2.5) {F};
\node[anchor=east,inner sep=2pt] at (0,-3.5) {xarray-4629};
\filldraw[fill=teal!16,draw=white,line width=.35pt] (0,-3) rectangle (1,-4);
\node[inner sep=0pt] at (0.5,-3.5) {P};
\filldraw[fill=teal!16,draw=white,line width=.35pt] (1,-3) rectangle (2,-4);
\node[inner sep=0pt] at (1.5,-3.5) {P};
\filldraw[fill=teal!16,draw=white,line width=.35pt] (2,-3) rectangle (3,-4);
\node[inner sep=0pt] at (2.5,-3.5) {P};
\filldraw[fill=teal!16,draw=white,line width=.35pt] (3,-3) rectangle (4,-4);
\node[inner sep=0pt] at (3.5,-3.5) {P};
\filldraw[fill=teal!16,draw=white,line width=.35pt] (4,-3) rectangle (5,-4);
\node[inner sep=0pt] at (4.5,-3.5) {P};
\filldraw[fill=teal!16,draw=white,line width=.35pt] (5,-3) rectangle (6,-4);
\node[inner sep=0pt] at (5.5,-3.5) {P};
\filldraw[fill=teal!16,draw=white,line width=.35pt] (6,-3) rectangle (7,-4);
\node[inner sep=0pt] at (6.5,-3.5) {P};
\filldraw[fill=teal!16,draw=white,line width=.35pt] (7,-3) rectangle (8,-4);
\node[inner sep=0pt] at (7.5,-3.5) {P};
\filldraw[fill=teal!16,draw=white,line width=.35pt] (8,-3) rectangle (9,-4);
\node[inner sep=0pt] at (8.5,-3.5) {P};
\filldraw[fill=teal!16,draw=white,line width=.35pt] (9,-3) rectangle (10,-4);
\node[inner sep=0pt] at (9.5,-3.5) {P};
\filldraw[fill=teal!16,draw=white,line width=.35pt] (10,-3) rectangle (11,-4);
\node[inner sep=0pt] at (10.5,-3.5) {P};
\filldraw[fill=teal!16,draw=white,line width=.35pt] (11,-3) rectangle (12,-4);
\node[inner sep=0pt] at (11.5,-3.5) {P};
\filldraw[fill=teal!16,draw=white,line width=.35pt] (12,-3) rectangle (13,-4);
\node[inner sep=0pt] at (12.5,-3.5) {P};
\filldraw[fill=teal!16,draw=white,line width=.35pt] (13,-3) rectangle (14,-4);
\node[inner sep=0pt] at (13.5,-3.5) {P};
\node[anchor=east,inner sep=2pt] at (0,-4.5) {xarray-6721};
\filldraw[fill=teal!16,draw=white,line width=.35pt] (0,-4) rectangle (1,-5);
\node[inner sep=0pt] at (0.5,-4.5) {P};
\filldraw[fill=teal!16,draw=white,line width=.35pt] (1,-4) rectangle (2,-5);
\node[inner sep=0pt] at (1.5,-4.5) {P};
\filldraw[fill=orange!28,draw=white,line width=.35pt] (2,-4) rectangle (3,-5);
\node[inner sep=0pt] at (2.5,-4.5) {F};
\filldraw[fill=teal!16,draw=white,line width=.35pt] (3,-4) rectangle (4,-5);
\node[inner sep=0pt] at (3.5,-4.5) {P};
\filldraw[fill=teal!16,draw=white,line width=.35pt] (4,-4) rectangle (5,-5);
\node[inner sep=0pt] at (4.5,-4.5) {P};
\filldraw[fill=teal!16,draw=white,line width=.35pt] (5,-4) rectangle (6,-5);
\node[inner sep=0pt] at (5.5,-4.5) {P};
\filldraw[fill=teal!16,draw=white,line width=.35pt] (6,-4) rectangle (7,-5);
\node[inner sep=0pt] at (6.5,-4.5) {P};
\filldraw[fill=orange!28,draw=white,line width=.35pt] (7,-4) rectangle (8,-5);
\node[inner sep=0pt] at (7.5,-4.5) {F};
\filldraw[fill=teal!16,draw=white,line width=.35pt] (8,-4) rectangle (9,-5);
\node[inner sep=0pt] at (8.5,-4.5) {P};
\filldraw[fill=teal!16,draw=white,line width=.35pt] (9,-4) rectangle (10,-5);
\node[inner sep=0pt] at (9.5,-4.5) {P};
\filldraw[fill=teal!16,draw=white,line width=.35pt] (10,-4) rectangle (11,-5);
\node[inner sep=0pt] at (10.5,-4.5) {P};
\filldraw[fill=teal!16,draw=white,line width=.35pt] (11,-4) rectangle (12,-5);
\node[inner sep=0pt] at (11.5,-4.5) {P};
\filldraw[fill=teal!16,draw=white,line width=.35pt] (12,-4) rectangle (13,-5);
\node[inner sep=0pt] at (12.5,-4.5) {P};
\filldraw[fill=teal!16,draw=white,line width=.35pt] (13,-4) rectangle (14,-5);
\node[inner sep=0pt] at (13.5,-4.5) {P};
\node[anchor=east,inner sep=2pt] at (0,-5.5) {pylint-6903};
\filldraw[fill=teal!16,draw=white,line width=.35pt] (0,-5) rectangle (1,-6);
\node[inner sep=0pt] at (0.5,-5.5) {P};
\filldraw[fill=teal!16,draw=white,line width=.35pt] (1,-5) rectangle (2,-6);
\node[inner sep=0pt] at (1.5,-5.5) {P};
\filldraw[fill=teal!16,draw=white,line width=.35pt] (2,-5) rectangle (3,-6);
\node[inner sep=0pt] at (2.5,-5.5) {P};
\filldraw[fill=teal!16,draw=white,line width=.35pt] (3,-5) rectangle (4,-6);
\node[inner sep=0pt] at (3.5,-5.5) {P};
\filldraw[fill=teal!16,draw=white,line width=.35pt] (4,-5) rectangle (5,-6);
\node[inner sep=0pt] at (4.5,-5.5) {P};
\filldraw[fill=teal!16,draw=white,line width=.35pt] (5,-5) rectangle (6,-6);
\node[inner sep=0pt] at (5.5,-5.5) {P};
\filldraw[fill=teal!16,draw=white,line width=.35pt] (6,-5) rectangle (7,-6);
\node[inner sep=0pt] at (6.5,-5.5) {P};
\filldraw[fill=teal!16,draw=white,line width=.35pt] (7,-5) rectangle (8,-6);
\node[inner sep=0pt] at (7.5,-5.5) {P};
\filldraw[fill=teal!16,draw=white,line width=.35pt] (8,-5) rectangle (9,-6);
\node[inner sep=0pt] at (8.5,-5.5) {P};
\filldraw[fill=teal!16,draw=white,line width=.35pt] (9,-5) rectangle (10,-6);
\node[inner sep=0pt] at (9.5,-5.5) {P};
\filldraw[fill=teal!16,draw=white,line width=.35pt] (10,-5) rectangle (11,-6);
\node[inner sep=0pt] at (10.5,-5.5) {P};
\filldraw[fill=teal!16,draw=white,line width=.35pt] (11,-5) rectangle (12,-6);
\node[inner sep=0pt] at (11.5,-5.5) {P};
\filldraw[fill=teal!16,draw=white,line width=.35pt] (12,-5) rectangle (13,-6);
\node[inner sep=0pt] at (12.5,-5.5) {P};
\filldraw[fill=teal!16,draw=white,line width=.35pt] (13,-5) rectangle (14,-6);
\node[inner sep=0pt] at (13.5,-5.5) {P};
\node[anchor=east,inner sep=2pt] at (0,-6.5) {pytest-10081};
\filldraw[fill=teal!16,draw=white,line width=.35pt] (0,-6) rectangle (1,-7);
\node[inner sep=0pt] at (0.5,-6.5) {P};
\filldraw[fill=orange!28,draw=white,line width=.35pt] (1,-6) rectangle (2,-7);
\node[inner sep=0pt] at (1.5,-6.5) {F};
\filldraw[fill=orange!28,draw=white,line width=.35pt] (2,-6) rectangle (3,-7);
\node[inner sep=0pt] at (2.5,-6.5) {F};
\filldraw[fill=orange!28,draw=white,line width=.35pt] (3,-6) rectangle (4,-7);
\node[inner sep=0pt] at (3.5,-6.5) {F};
\filldraw[fill=orange!28,draw=white,line width=.35pt] (4,-6) rectangle (5,-7);
\node[inner sep=0pt] at (4.5,-6.5) {F};
\filldraw[fill=orange!28,draw=white,line width=.35pt] (5,-6) rectangle (6,-7);
\node[inner sep=0pt] at (5.5,-6.5) {F};
\filldraw[fill=orange!28,draw=white,line width=.35pt] (6,-6) rectangle (7,-7);
\node[inner sep=0pt] at (6.5,-6.5) {F};
\filldraw[fill=orange!28,draw=white,line width=.35pt] (7,-6) rectangle (8,-7);
\node[inner sep=0pt] at (7.5,-6.5) {F};
\filldraw[fill=orange!28,draw=white,line width=.35pt] (8,-6) rectangle (9,-7);
\node[inner sep=0pt] at (8.5,-6.5) {F};
\filldraw[fill=orange!28,draw=white,line width=.35pt] (9,-6) rectangle (10,-7);
\node[inner sep=0pt] at (9.5,-6.5) {F};
\filldraw[fill=orange!28,draw=white,line width=.35pt] (10,-6) rectangle (11,-7);
\node[inner sep=0pt] at (10.5,-6.5) {F};
\filldraw[fill=orange!28,draw=white,line width=.35pt] (11,-6) rectangle (12,-7);
\node[inner sep=0pt] at (11.5,-6.5) {F};
\filldraw[fill=orange!28,draw=white,line width=.35pt] (12,-6) rectangle (13,-7);
\node[inner sep=0pt] at (12.5,-6.5) {F};
\filldraw[fill=orange!28,draw=white,line width=.35pt] (13,-6) rectangle (14,-7);
\node[inner sep=0pt] at (13.5,-6.5) {F};
\node[anchor=east,inner sep=2pt] at (0,-7.5) {pytest-5631};
\filldraw[fill=teal!16,draw=white,line width=.35pt] (0,-7) rectangle (1,-8);
\node[inner sep=0pt] at (0.5,-7.5) {P};
\filldraw[fill=teal!16,draw=white,line width=.35pt] (1,-7) rectangle (2,-8);
\node[inner sep=0pt] at (1.5,-7.5) {P};
\filldraw[fill=teal!16,draw=white,line width=.35pt] (2,-7) rectangle (3,-8);
\node[inner sep=0pt] at (2.5,-7.5) {P};
\filldraw[fill=teal!16,draw=white,line width=.35pt] (3,-7) rectangle (4,-8);
\node[inner sep=0pt] at (3.5,-7.5) {P};
\filldraw[fill=teal!16,draw=white,line width=.35pt] (4,-7) rectangle (5,-8);
\node[inner sep=0pt] at (4.5,-7.5) {P};
\filldraw[fill=teal!16,draw=white,line width=.35pt] (5,-7) rectangle (6,-8);
\node[inner sep=0pt] at (5.5,-7.5) {P};
\filldraw[fill=teal!16,draw=white,line width=.35pt] (6,-7) rectangle (7,-8);
\node[inner sep=0pt] at (6.5,-7.5) {P};
\filldraw[fill=orange!28,draw=white,line width=.35pt] (7,-7) rectangle (8,-8);
\node[inner sep=0pt] at (7.5,-7.5) {F};
\filldraw[fill=teal!16,draw=white,line width=.35pt] (8,-7) rectangle (9,-8);
\node[inner sep=0pt] at (8.5,-7.5) {P};
\filldraw[fill=teal!16,draw=white,line width=.35pt] (9,-7) rectangle (10,-8);
\node[inner sep=0pt] at (9.5,-7.5) {P};
\filldraw[fill=teal!16,draw=white,line width=.35pt] (10,-7) rectangle (11,-8);
\node[inner sep=0pt] at (10.5,-7.5) {P};
\filldraw[fill=teal!16,draw=white,line width=.35pt] (11,-7) rectangle (12,-8);
\node[inner sep=0pt] at (11.5,-7.5) {P};
\filldraw[fill=orange!28,draw=white,line width=.35pt] (12,-7) rectangle (13,-8);
\node[inner sep=0pt] at (12.5,-7.5) {F};
\filldraw[fill=orange!28,draw=white,line width=.35pt] (13,-7) rectangle (14,-8);
\node[inner sep=0pt] at (13.5,-7.5) {F};
\node[anchor=east,inner sep=2pt] at (0,-8.5) {pytest-5809};
\filldraw[fill=teal!16,draw=white,line width=.35pt] (0,-8) rectangle (1,-9);
\node[inner sep=0pt] at (0.5,-8.5) {P};
\filldraw[fill=teal!16,draw=white,line width=.35pt] (1,-8) rectangle (2,-9);
\node[inner sep=0pt] at (1.5,-8.5) {P};
\filldraw[fill=teal!16,draw=white,line width=.35pt] (2,-8) rectangle (3,-9);
\node[inner sep=0pt] at (2.5,-8.5) {P};
\filldraw[fill=teal!16,draw=white,line width=.35pt] (3,-8) rectangle (4,-9);
\node[inner sep=0pt] at (3.5,-8.5) {P};
\filldraw[fill=teal!16,draw=white,line width=.35pt] (4,-8) rectangle (5,-9);
\node[inner sep=0pt] at (4.5,-8.5) {P};
\filldraw[fill=teal!16,draw=white,line width=.35pt] (5,-8) rectangle (6,-9);
\node[inner sep=0pt] at (5.5,-8.5) {P};
\filldraw[fill=orange!28,draw=white,line width=.35pt] (6,-8) rectangle (7,-9);
\node[inner sep=0pt] at (6.5,-8.5) {F};
\filldraw[fill=teal!16,draw=white,line width=.35pt] (7,-8) rectangle (8,-9);
\node[inner sep=0pt] at (7.5,-8.5) {P};
\filldraw[fill=orange!28,draw=white,line width=.35pt] (8,-8) rectangle (9,-9);
\node[inner sep=0pt] at (8.5,-8.5) {F};
\filldraw[fill=teal!16,draw=white,line width=.35pt] (9,-8) rectangle (10,-9);
\node[inner sep=0pt] at (9.5,-8.5) {P};
\filldraw[fill=teal!16,draw=white,line width=.35pt] (10,-8) rectangle (11,-9);
\node[inner sep=0pt] at (10.5,-8.5) {P};
\filldraw[fill=teal!16,draw=white,line width=.35pt] (11,-8) rectangle (12,-9);
\node[inner sep=0pt] at (11.5,-8.5) {P};
\filldraw[fill=teal!16,draw=white,line width=.35pt] (12,-8) rectangle (13,-9);
\node[inner sep=0pt] at (12.5,-8.5) {P};
\filldraw[fill=teal!16,draw=white,line width=.35pt] (13,-8) rectangle (14,-9);
\node[inner sep=0pt] at (13.5,-8.5) {P};
\node[anchor=east,inner sep=2pt] at (0,-9.5) {pytest-6202};
\filldraw[fill=teal!16,draw=white,line width=.35pt] (0,-9) rectangle (1,-10);
\node[inner sep=0pt] at (0.5,-9.5) {P};
\filldraw[fill=teal!16,draw=white,line width=.35pt] (1,-9) rectangle (2,-10);
\node[inner sep=0pt] at (1.5,-9.5) {P};
\filldraw[fill=teal!16,draw=white,line width=.35pt] (2,-9) rectangle (3,-10);
\node[inner sep=0pt] at (2.5,-9.5) {P};
\filldraw[fill=teal!16,draw=white,line width=.35pt] (3,-9) rectangle (4,-10);
\node[inner sep=0pt] at (3.5,-9.5) {P};
\filldraw[fill=teal!16,draw=white,line width=.35pt] (4,-9) rectangle (5,-10);
\node[inner sep=0pt] at (4.5,-9.5) {P};
\filldraw[fill=orange!28,draw=white,line width=.35pt] (5,-9) rectangle (6,-10);
\node[inner sep=0pt] at (5.5,-9.5) {F};
\filldraw[fill=orange!28,draw=white,line width=.35pt] (6,-9) rectangle (7,-10);
\node[inner sep=0pt] at (6.5,-9.5) {F};
\filldraw[fill=orange!28,draw=white,line width=.35pt] (7,-9) rectangle (8,-10);
\node[inner sep=0pt] at (7.5,-9.5) {F};
\filldraw[fill=orange!28,draw=white,line width=.35pt] (8,-9) rectangle (9,-10);
\node[inner sep=0pt] at (8.5,-9.5) {F};
\filldraw[fill=orange!28,draw=white,line width=.35pt] (9,-9) rectangle (10,-10);
\node[inner sep=0pt] at (9.5,-9.5) {F};
\filldraw[fill=orange!28,draw=white,line width=.35pt] (10,-9) rectangle (11,-10);
\node[inner sep=0pt] at (10.5,-9.5) {F};
\filldraw[fill=orange!28,draw=white,line width=.35pt] (11,-9) rectangle (12,-10);
\node[inner sep=0pt] at (11.5,-9.5) {F};
\filldraw[fill=orange!28,draw=white,line width=.35pt] (12,-9) rectangle (13,-10);
\node[inner sep=0pt] at (12.5,-9.5) {F};
\filldraw[fill=orange!28,draw=white,line width=.35pt] (13,-9) rectangle (14,-10);
\node[inner sep=0pt] at (13.5,-9.5) {F};
\node[anchor=east,inner sep=2pt] at (0,-10.5) {pytest-7205};
\filldraw[fill=teal!16,draw=white,line width=.35pt] (0,-10) rectangle (1,-11);
\node[inner sep=0pt] at (0.5,-10.5) {P};
\filldraw[fill=teal!16,draw=white,line width=.35pt] (1,-10) rectangle (2,-11);
\node[inner sep=0pt] at (1.5,-10.5) {P};
\filldraw[fill=teal!16,draw=white,line width=.35pt] (2,-10) rectangle (3,-11);
\node[inner sep=0pt] at (2.5,-10.5) {P};
\filldraw[fill=teal!16,draw=white,line width=.35pt] (3,-10) rectangle (4,-11);
\node[inner sep=0pt] at (3.5,-10.5) {P};
\filldraw[fill=orange!28,draw=white,line width=.35pt] (4,-10) rectangle (5,-11);
\node[inner sep=0pt] at (4.5,-10.5) {F};
\filldraw[fill=teal!16,draw=white,line width=.35pt] (5,-10) rectangle (6,-11);
\node[inner sep=0pt] at (5.5,-10.5) {P};
\filldraw[fill=teal!16,draw=white,line width=.35pt] (6,-10) rectangle (7,-11);
\node[inner sep=0pt] at (6.5,-10.5) {P};
\filldraw[fill=teal!16,draw=white,line width=.35pt] (7,-10) rectangle (8,-11);
\node[inner sep=0pt] at (7.5,-10.5) {P};
\filldraw[fill=teal!16,draw=white,line width=.35pt] (8,-10) rectangle (9,-11);
\node[inner sep=0pt] at (8.5,-10.5) {P};
\filldraw[fill=orange!28,draw=white,line width=.35pt] (9,-10) rectangle (10,-11);
\node[inner sep=0pt] at (9.5,-10.5) {F};
\filldraw[fill=orange!28,draw=white,line width=.35pt] (10,-10) rectangle (11,-11);
\node[inner sep=0pt] at (10.5,-10.5) {F};
\filldraw[fill=orange!28,draw=white,line width=.35pt] (11,-10) rectangle (12,-11);
\node[inner sep=0pt] at (11.5,-10.5) {F};
\filldraw[fill=orange!28,draw=white,line width=.35pt] (12,-10) rectangle (13,-11);
\node[inner sep=0pt] at (12.5,-10.5) {F};
\filldraw[fill=orange!28,draw=white,line width=.35pt] (13,-10) rectangle (14,-11);
\node[inner sep=0pt] at (13.5,-10.5) {F};
\node[anchor=east,inner sep=2pt] at (0,-11.5) {pytest-7432};
\filldraw[fill=teal!16,draw=white,line width=.35pt] (0,-11) rectangle (1,-12);
\node[inner sep=0pt] at (0.5,-11.5) {P};
\filldraw[fill=teal!16,draw=white,line width=.35pt] (1,-11) rectangle (2,-12);
\node[inner sep=0pt] at (1.5,-11.5) {P};
\filldraw[fill=teal!16,draw=white,line width=.35pt] (2,-11) rectangle (3,-12);
\node[inner sep=0pt] at (2.5,-11.5) {P};
\filldraw[fill=teal!16,draw=white,line width=.35pt] (3,-11) rectangle (4,-12);
\node[inner sep=0pt] at (3.5,-11.5) {P};
\filldraw[fill=teal!16,draw=white,line width=.35pt] (4,-11) rectangle (5,-12);
\node[inner sep=0pt] at (4.5,-11.5) {P};
\filldraw[fill=teal!16,draw=white,line width=.35pt] (5,-11) rectangle (6,-12);
\node[inner sep=0pt] at (5.5,-11.5) {P};
\filldraw[fill=orange!28,draw=white,line width=.35pt] (6,-11) rectangle (7,-12);
\node[inner sep=0pt] at (6.5,-11.5) {F};
\filldraw[fill=orange!28,draw=white,line width=.35pt] (7,-11) rectangle (8,-12);
\node[inner sep=0pt] at (7.5,-11.5) {F};
\filldraw[fill=teal!16,draw=white,line width=.35pt] (8,-11) rectangle (9,-12);
\node[inner sep=0pt] at (8.5,-11.5) {P};
\filldraw[fill=orange!28,draw=white,line width=.35pt] (9,-11) rectangle (10,-12);
\node[inner sep=0pt] at (9.5,-11.5) {F};
\filldraw[fill=orange!28,draw=white,line width=.35pt] (10,-11) rectangle (11,-12);
\node[inner sep=0pt] at (10.5,-11.5) {F};
\filldraw[fill=teal!16,draw=white,line width=.35pt] (11,-11) rectangle (12,-12);
\node[inner sep=0pt] at (11.5,-11.5) {P};
\filldraw[fill=orange!28,draw=white,line width=.35pt] (12,-11) rectangle (13,-12);
\node[inner sep=0pt] at (12.5,-11.5) {F};
\filldraw[fill=orange!28,draw=white,line width=.35pt] (13,-11) rectangle (14,-12);
\node[inner sep=0pt] at (13.5,-11.5) {F};
\node[anchor=east,inner sep=2pt] at (0,-12.5) {pytest-7490};
\filldraw[fill=teal!16,draw=white,line width=.35pt] (0,-12) rectangle (1,-13);
\node[inner sep=0pt] at (0.5,-12.5) {P};
\filldraw[fill=orange!28,draw=white,line width=.35pt] (1,-12) rectangle (2,-13);
\node[inner sep=0pt] at (1.5,-12.5) {F};
\filldraw[fill=teal!16,draw=white,line width=.35pt] (2,-12) rectangle (3,-13);
\node[inner sep=0pt] at (2.5,-12.5) {P};
\filldraw[fill=teal!16,draw=white,line width=.35pt] (3,-12) rectangle (4,-13);
\node[inner sep=0pt] at (3.5,-12.5) {P};
\filldraw[fill=orange!28,draw=white,line width=.35pt] (4,-12) rectangle (5,-13);
\node[inner sep=0pt] at (4.5,-12.5) {F};
\filldraw[fill=orange!28,draw=white,line width=.35pt] (5,-12) rectangle (6,-13);
\node[inner sep=0pt] at (5.5,-12.5) {F};
\filldraw[fill=orange!28,draw=white,line width=.35pt] (6,-12) rectangle (7,-13);
\node[inner sep=0pt] at (6.5,-12.5) {F};
\filldraw[fill=orange!28,draw=white,line width=.35pt] (7,-12) rectangle (8,-13);
\node[inner sep=0pt] at (7.5,-12.5) {F};
\filldraw[fill=orange!28,draw=white,line width=.35pt] (8,-12) rectangle (9,-13);
\node[inner sep=0pt] at (8.5,-12.5) {F};
\filldraw[fill=orange!28,draw=white,line width=.35pt] (9,-12) rectangle (10,-13);
\node[inner sep=0pt] at (9.5,-12.5) {F};
\filldraw[fill=orange!28,draw=white,line width=.35pt] (10,-12) rectangle (11,-13);
\node[inner sep=0pt] at (10.5,-12.5) {F};
\filldraw[fill=orange!28,draw=white,line width=.35pt] (11,-12) rectangle (12,-13);
\node[inner sep=0pt] at (11.5,-12.5) {F};
\filldraw[fill=orange!28,draw=white,line width=.35pt] (12,-12) rectangle (13,-13);
\node[inner sep=0pt] at (12.5,-12.5) {F};
\filldraw[fill=orange!28,draw=white,line width=.35pt] (13,-12) rectangle (14,-13);
\node[inner sep=0pt] at (13.5,-12.5) {F};
\node[anchor=east,inner sep=2pt] at (0,-13.5) {pytest-7521};
\filldraw[fill=teal!16,draw=white,line width=.35pt] (0,-13) rectangle (1,-14);
\node[inner sep=0pt] at (0.5,-13.5) {P};
\filldraw[fill=teal!16,draw=white,line width=.35pt] (1,-13) rectangle (2,-14);
\node[inner sep=0pt] at (1.5,-13.5) {P};
\filldraw[fill=teal!16,draw=white,line width=.35pt] (2,-13) rectangle (3,-14);
\node[inner sep=0pt] at (2.5,-13.5) {P};
\filldraw[fill=teal!16,draw=white,line width=.35pt] (3,-13) rectangle (4,-14);
\node[inner sep=0pt] at (3.5,-13.5) {P};
\filldraw[fill=orange!28,draw=white,line width=.35pt] (4,-13) rectangle (5,-14);
\node[inner sep=0pt] at (4.5,-13.5) {F};
\filldraw[fill=orange!28,draw=white,line width=.35pt] (5,-13) rectangle (6,-14);
\node[inner sep=0pt] at (5.5,-13.5) {F};
\filldraw[fill=orange!28,draw=white,line width=.35pt] (6,-13) rectangle (7,-14);
\node[inner sep=0pt] at (6.5,-13.5) {F};
\filldraw[fill=orange!28,draw=white,line width=.35pt] (7,-13) rectangle (8,-14);
\node[inner sep=0pt] at (7.5,-13.5) {F};
\filldraw[fill=orange!28,draw=white,line width=.35pt] (8,-13) rectangle (9,-14);
\node[inner sep=0pt] at (8.5,-13.5) {F};
\filldraw[fill=orange!28,draw=white,line width=.35pt] (9,-13) rectangle (10,-14);
\node[inner sep=0pt] at (9.5,-13.5) {F};
\filldraw[fill=orange!28,draw=white,line width=.35pt] (10,-13) rectangle (11,-14);
\node[inner sep=0pt] at (10.5,-13.5) {F};
\filldraw[fill=orange!28,draw=white,line width=.35pt] (11,-13) rectangle (12,-14);
\node[inner sep=0pt] at (11.5,-13.5) {F};
\filldraw[fill=orange!28,draw=white,line width=.35pt] (12,-13) rectangle (13,-14);
\node[inner sep=0pt] at (12.5,-13.5) {F};
\filldraw[fill=orange!28,draw=white,line width=.35pt] (13,-13) rectangle (14,-14);
\node[inner sep=0pt] at (13.5,-13.5) {F};
\node[anchor=east,inner sep=2pt] at (0,-14.5) {pytest-7571};
\filldraw[fill=teal!16,draw=white,line width=.35pt] (0,-14) rectangle (1,-15);
\node[inner sep=0pt] at (0.5,-14.5) {P};
\filldraw[fill=orange!28,draw=white,line width=.35pt] (1,-14) rectangle (2,-15);
\node[inner sep=0pt] at (1.5,-14.5) {F};
\filldraw[fill=teal!16,draw=white,line width=.35pt] (2,-14) rectangle (3,-15);
\node[inner sep=0pt] at (2.5,-14.5) {P};
\filldraw[fill=teal!16,draw=white,line width=.35pt] (3,-14) rectangle (4,-15);
\node[inner sep=0pt] at (3.5,-14.5) {P};
\filldraw[fill=orange!28,draw=white,line width=.35pt] (4,-14) rectangle (5,-15);
\node[inner sep=0pt] at (4.5,-14.5) {F};
\filldraw[fill=orange!28,draw=white,line width=.35pt] (5,-14) rectangle (6,-15);
\node[inner sep=0pt] at (5.5,-14.5) {F};
\filldraw[fill=orange!28,draw=white,line width=.35pt] (6,-14) rectangle (7,-15);
\node[inner sep=0pt] at (6.5,-14.5) {F};
\filldraw[fill=orange!28,draw=white,line width=.35pt] (7,-14) rectangle (8,-15);
\node[inner sep=0pt] at (7.5,-14.5) {F};
\filldraw[fill=teal!16,draw=white,line width=.35pt] (8,-14) rectangle (9,-15);
\node[inner sep=0pt] at (8.5,-14.5) {P};
\filldraw[fill=orange!28,draw=white,line width=.35pt] (9,-14) rectangle (10,-15);
\node[inner sep=0pt] at (9.5,-14.5) {F};
\filldraw[fill=orange!28,draw=white,line width=.35pt] (10,-14) rectangle (11,-15);
\node[inner sep=0pt] at (10.5,-14.5) {F};
\filldraw[fill=orange!28,draw=white,line width=.35pt] (11,-14) rectangle (12,-15);
\node[inner sep=0pt] at (11.5,-14.5) {F};
\filldraw[fill=orange!28,draw=white,line width=.35pt] (12,-14) rectangle (13,-15);
\node[inner sep=0pt] at (12.5,-14.5) {F};
\filldraw[fill=orange!28,draw=white,line width=.35pt] (13,-14) rectangle (14,-15);
\node[inner sep=0pt] at (13.5,-14.5) {F};
\node[anchor=east,inner sep=2pt] at (0,-15.5) {pytest-7982};
\filldraw[fill=teal!16,draw=white,line width=.35pt] (0,-15) rectangle (1,-16);
\node[inner sep=0pt] at (0.5,-15.5) {P};
\filldraw[fill=teal!16,draw=white,line width=.35pt] (1,-15) rectangle (2,-16);
\node[inner sep=0pt] at (1.5,-15.5) {P};
\filldraw[fill=teal!16,draw=white,line width=.35pt] (2,-15) rectangle (3,-16);
\node[inner sep=0pt] at (2.5,-15.5) {P};
\filldraw[fill=teal!16,draw=white,line width=.35pt] (3,-15) rectangle (4,-16);
\node[inner sep=0pt] at (3.5,-15.5) {P};
\filldraw[fill=orange!28,draw=white,line width=.35pt] (4,-15) rectangle (5,-16);
\node[inner sep=0pt] at (4.5,-15.5) {F};
\filldraw[fill=orange!28,draw=white,line width=.35pt] (5,-15) rectangle (6,-16);
\node[inner sep=0pt] at (5.5,-15.5) {F};
\filldraw[fill=orange!28,draw=white,line width=.35pt] (6,-15) rectangle (7,-16);
\node[inner sep=0pt] at (6.5,-15.5) {F};
\filldraw[fill=orange!28,draw=white,line width=.35pt] (7,-15) rectangle (8,-16);
\node[inner sep=0pt] at (7.5,-15.5) {F};
\filldraw[fill=teal!16,draw=white,line width=.35pt] (8,-15) rectangle (9,-16);
\node[inner sep=0pt] at (8.5,-15.5) {P};
\filldraw[fill=orange!28,draw=white,line width=.35pt] (9,-15) rectangle (10,-16);
\node[inner sep=0pt] at (9.5,-15.5) {F};
\filldraw[fill=orange!28,draw=white,line width=.35pt] (10,-15) rectangle (11,-16);
\node[inner sep=0pt] at (10.5,-15.5) {F};
\filldraw[fill=orange!28,draw=white,line width=.35pt] (11,-15) rectangle (12,-16);
\node[inner sep=0pt] at (11.5,-15.5) {F};
\filldraw[fill=orange!28,draw=white,line width=.35pt] (12,-15) rectangle (13,-16);
\node[inner sep=0pt] at (12.5,-15.5) {F};
\filldraw[fill=orange!28,draw=white,line width=.35pt] (13,-15) rectangle (14,-16);
\node[inner sep=0pt] at (13.5,-15.5) {F};
\node[anchor=east,inner sep=2pt] at (0,-16.5) {pytest-8399};
\filldraw[fill=teal!16,draw=white,line width=.35pt] (0,-16) rectangle (1,-17);
\node[inner sep=0pt] at (0.5,-16.5) {P};
\filldraw[fill=teal!16,draw=white,line width=.35pt] (1,-16) rectangle (2,-17);
\node[inner sep=0pt] at (1.5,-16.5) {P};
\filldraw[fill=teal!16,draw=white,line width=.35pt] (2,-16) rectangle (3,-17);
\node[inner sep=0pt] at (2.5,-16.5) {P};
\filldraw[fill=teal!16,draw=white,line width=.35pt] (3,-16) rectangle (4,-17);
\node[inner sep=0pt] at (3.5,-16.5) {P};
\filldraw[fill=orange!28,draw=white,line width=.35pt] (4,-16) rectangle (5,-17);
\node[inner sep=0pt] at (4.5,-16.5) {F};
\filldraw[fill=orange!28,draw=white,line width=.35pt] (5,-16) rectangle (6,-17);
\node[inner sep=0pt] at (5.5,-16.5) {F};
\filldraw[fill=orange!28,draw=white,line width=.35pt] (6,-16) rectangle (7,-17);
\node[inner sep=0pt] at (6.5,-16.5) {F};
\filldraw[fill=orange!28,draw=white,line width=.35pt] (7,-16) rectangle (8,-17);
\node[inner sep=0pt] at (7.5,-16.5) {F};
\filldraw[fill=orange!28,draw=white,line width=.35pt] (8,-16) rectangle (9,-17);
\node[inner sep=0pt] at (8.5,-16.5) {F};
\filldraw[fill=orange!28,draw=white,line width=.35pt] (9,-16) rectangle (10,-17);
\node[inner sep=0pt] at (9.5,-16.5) {F};
\filldraw[fill=orange!28,draw=white,line width=.35pt] (10,-16) rectangle (11,-17);
\node[inner sep=0pt] at (10.5,-16.5) {F};
\filldraw[fill=orange!28,draw=white,line width=.35pt] (11,-16) rectangle (12,-17);
\node[inner sep=0pt] at (11.5,-16.5) {F};
\filldraw[fill=orange!28,draw=white,line width=.35pt] (12,-16) rectangle (13,-17);
\node[inner sep=0pt] at (12.5,-16.5) {F};
\filldraw[fill=orange!28,draw=white,line width=.35pt] (13,-16) rectangle (14,-17);
\node[inner sep=0pt] at (13.5,-16.5) {F};
\node[anchor=east,inner sep=2pt] at (0,-17.5) {sphinx-9230};
\filldraw[fill=orange!28,draw=white,line width=.35pt] (0,-17) rectangle (1,-18);
\node[inner sep=0pt] at (0.5,-17.5) {F};
\filldraw[fill=orange!28,draw=white,line width=.35pt] (1,-17) rectangle (2,-18);
\node[inner sep=0pt] at (1.5,-17.5) {F};
\filldraw[fill=orange!28,draw=white,line width=.35pt] (2,-17) rectangle (3,-18);
\node[inner sep=0pt] at (2.5,-17.5) {F};
\filldraw[fill=orange!28,draw=white,line width=.35pt] (3,-17) rectangle (4,-18);
\node[inner sep=0pt] at (3.5,-17.5) {F};
\filldraw[fill=teal!16,draw=white,line width=.35pt] (4,-17) rectangle (5,-18);
\node[inner sep=0pt] at (4.5,-17.5) {P};
\filldraw[fill=teal!16,draw=white,line width=.35pt] (5,-17) rectangle (6,-18);
\node[inner sep=0pt] at (5.5,-17.5) {P};
\filldraw[fill=teal!16,draw=white,line width=.35pt] (6,-17) rectangle (7,-18);
\node[inner sep=0pt] at (6.5,-17.5) {P};
\filldraw[fill=teal!16,draw=white,line width=.35pt] (7,-17) rectangle (8,-18);
\node[inner sep=0pt] at (7.5,-17.5) {P};
\filldraw[fill=orange!28,draw=white,line width=.35pt] (8,-17) rectangle (9,-18);
\node[inner sep=0pt] at (8.5,-17.5) {F};
\filldraw[fill=orange!28,draw=white,line width=.35pt] (9,-17) rectangle (10,-18);
\node[inner sep=0pt] at (9.5,-17.5) {F};
\filldraw[fill=orange!28,draw=white,line width=.35pt] (10,-17) rectangle (11,-18);
\node[inner sep=0pt] at (10.5,-17.5) {F};
\filldraw[fill=orange!28,draw=white,line width=.35pt] (11,-17) rectangle (12,-18);
\node[inner sep=0pt] at (11.5,-17.5) {F};
\filldraw[fill=orange!28,draw=white,line width=.35pt] (12,-17) rectangle (13,-18);
\node[inner sep=0pt] at (12.5,-17.5) {F};
\filldraw[fill=teal!16,draw=white,line width=.35pt] (13,-17) rectangle (14,-18);
\node[inner sep=0pt] at (13.5,-17.5) {P};
\node[anchor=east,inner sep=2pt] at (0,-18.5) {sphinx-9591};
\filldraw[fill=teal!16,draw=white,line width=.35pt] (0,-18) rectangle (1,-19);
\node[inner sep=0pt] at (0.5,-18.5) {P};
\filldraw[fill=teal!16,draw=white,line width=.35pt] (1,-18) rectangle (2,-19);
\node[inner sep=0pt] at (1.5,-18.5) {P};
\filldraw[fill=teal!16,draw=white,line width=.35pt] (2,-18) rectangle (3,-19);
\node[inner sep=0pt] at (2.5,-18.5) {P};
\filldraw[fill=orange!28,draw=white,line width=.35pt] (3,-18) rectangle (4,-19);
\node[inner sep=0pt] at (3.5,-18.5) {F};
\filldraw[fill=teal!16,draw=white,line width=.35pt] (4,-18) rectangle (5,-19);
\node[inner sep=0pt] at (4.5,-18.5) {P};
\filldraw[fill=teal!16,draw=white,line width=.35pt] (5,-18) rectangle (6,-19);
\node[inner sep=0pt] at (5.5,-18.5) {P};
\filldraw[fill=orange!28,draw=white,line width=.35pt] (6,-18) rectangle (7,-19);
\node[inner sep=0pt] at (6.5,-18.5) {F};
\filldraw[fill=teal!16,draw=white,line width=.35pt] (7,-18) rectangle (8,-19);
\node[inner sep=0pt] at (7.5,-18.5) {P};
\filldraw[fill=orange!28,draw=white,line width=.35pt] (8,-18) rectangle (9,-19);
\node[inner sep=0pt] at (8.5,-18.5) {F};
\filldraw[fill=teal!16,draw=white,line width=.35pt] (9,-18) rectangle (10,-19);
\node[inner sep=0pt] at (9.5,-18.5) {P};
\filldraw[fill=orange!28,draw=white,line width=.35pt] (10,-18) rectangle (11,-19);
\node[inner sep=0pt] at (10.5,-18.5) {F};
\filldraw[fill=orange!28,draw=white,line width=.35pt] (11,-18) rectangle (12,-19);
\node[inner sep=0pt] at (11.5,-18.5) {F};
\filldraw[fill=teal!16,draw=white,line width=.35pt] (12,-18) rectangle (13,-19);
\node[inner sep=0pt] at (12.5,-18.5) {P};
\filldraw[fill=teal!16,draw=white,line width=.35pt] (13,-18) rectangle (14,-19);
\node[inner sep=0pt] at (13.5,-18.5) {P};
\node[anchor=east,inner sep=2pt] at (0,-19.5) {sympy-12481};
\filldraw[fill=teal!16,draw=white,line width=.35pt] (0,-19) rectangle (1,-20);
\node[inner sep=0pt] at (0.5,-19.5) {P};
\filldraw[fill=teal!16,draw=white,line width=.35pt] (1,-19) rectangle (2,-20);
\node[inner sep=0pt] at (1.5,-19.5) {P};
\filldraw[fill=orange!28,draw=white,line width=.35pt] (2,-19) rectangle (3,-20);
\node[inner sep=0pt] at (2.5,-19.5) {F};
\filldraw[fill=orange!28,draw=white,line width=.35pt] (3,-19) rectangle (4,-20);
\node[inner sep=0pt] at (3.5,-19.5) {F};
\filldraw[fill=teal!16,draw=white,line width=.35pt] (4,-19) rectangle (5,-20);
\node[inner sep=0pt] at (4.5,-19.5) {P};
\filldraw[fill=orange!28,draw=white,line width=.35pt] (5,-19) rectangle (6,-20);
\node[inner sep=0pt] at (5.5,-19.5) {F};
\filldraw[fill=teal!16,draw=white,line width=.35pt] (6,-19) rectangle (7,-20);
\node[inner sep=0pt] at (6.5,-19.5) {P};
\filldraw[fill=orange!28,draw=white,line width=.35pt] (7,-19) rectangle (8,-20);
\node[inner sep=0pt] at (7.5,-19.5) {F};
\filldraw[fill=orange!28,draw=white,line width=.35pt] (8,-19) rectangle (9,-20);
\node[inner sep=0pt] at (8.5,-19.5) {F};
\filldraw[fill=teal!16,draw=white,line width=.35pt] (9,-19) rectangle (10,-20);
\node[inner sep=0pt] at (9.5,-19.5) {P};
\filldraw[fill=teal!16,draw=white,line width=.35pt] (10,-19) rectangle (11,-20);
\node[inner sep=0pt] at (10.5,-19.5) {P};
\filldraw[fill=orange!28,draw=white,line width=.35pt] (11,-19) rectangle (12,-20);
\node[inner sep=0pt] at (11.5,-19.5) {F};
\filldraw[fill=orange!28,draw=white,line width=.35pt] (12,-19) rectangle (13,-20);
\node[inner sep=0pt] at (12.5,-19.5) {F};
\filldraw[fill=orange!28,draw=white,line width=.35pt] (13,-19) rectangle (14,-20);
\node[inner sep=0pt] at (13.5,-19.5) {F};
\node[anchor=east,inner sep=2pt] at (0,-20.5) {sympy-13480};
\filldraw[fill=teal!16,draw=white,line width=.35pt] (0,-20) rectangle (1,-21);
\node[inner sep=0pt] at (0.5,-20.5) {P};
\filldraw[fill=teal!16,draw=white,line width=.35pt] (1,-20) rectangle (2,-21);
\node[inner sep=0pt] at (1.5,-20.5) {P};
\filldraw[fill=teal!16,draw=white,line width=.35pt] (2,-20) rectangle (3,-21);
\node[inner sep=0pt] at (2.5,-20.5) {P};
\filldraw[fill=teal!16,draw=white,line width=.35pt] (3,-20) rectangle (4,-21);
\node[inner sep=0pt] at (3.5,-20.5) {P};
\filldraw[fill=teal!16,draw=white,line width=.35pt] (4,-20) rectangle (5,-21);
\node[inner sep=0pt] at (4.5,-20.5) {P};
\filldraw[fill=teal!16,draw=white,line width=.35pt] (5,-20) rectangle (6,-21);
\node[inner sep=0pt] at (5.5,-20.5) {P};
\filldraw[fill=teal!16,draw=white,line width=.35pt] (6,-20) rectangle (7,-21);
\node[inner sep=0pt] at (6.5,-20.5) {P};
\filldraw[fill=teal!16,draw=white,line width=.35pt] (7,-20) rectangle (8,-21);
\node[inner sep=0pt] at (7.5,-20.5) {P};
\filldraw[fill=teal!16,draw=white,line width=.35pt] (8,-20) rectangle (9,-21);
\node[inner sep=0pt] at (8.5,-20.5) {P};
\filldraw[fill=teal!16,draw=white,line width=.35pt] (9,-20) rectangle (10,-21);
\node[inner sep=0pt] at (9.5,-20.5) {P};
\filldraw[fill=teal!16,draw=white,line width=.35pt] (10,-20) rectangle (11,-21);
\node[inner sep=0pt] at (10.5,-20.5) {P};
\filldraw[fill=teal!16,draw=white,line width=.35pt] (11,-20) rectangle (12,-21);
\node[inner sep=0pt] at (11.5,-20.5) {P};
\filldraw[fill=teal!16,draw=white,line width=.35pt] (12,-20) rectangle (13,-21);
\node[inner sep=0pt] at (12.5,-20.5) {P};
\filldraw[fill=teal!16,draw=white,line width=.35pt] (13,-20) rectangle (14,-21);
\node[inner sep=0pt] at (13.5,-20.5) {P};
\node[anchor=east,inner sep=2pt] at (0,-21.5) {sympy-13551};
\filldraw[fill=teal!16,draw=white,line width=.35pt] (0,-21) rectangle (1,-22);
\node[inner sep=0pt] at (0.5,-21.5) {P};
\filldraw[fill=teal!16,draw=white,line width=.35pt] (1,-21) rectangle (2,-22);
\node[inner sep=0pt] at (1.5,-21.5) {P};
\filldraw[fill=teal!16,draw=white,line width=.35pt] (2,-21) rectangle (3,-22);
\node[inner sep=0pt] at (2.5,-21.5) {P};
\filldraw[fill=orange!28,draw=white,line width=.35pt] (3,-21) rectangle (4,-22);
\node[inner sep=0pt] at (3.5,-21.5) {F};
\filldraw[fill=teal!16,draw=white,line width=.35pt] (4,-21) rectangle (5,-22);
\node[inner sep=0pt] at (4.5,-21.5) {P};
\filldraw[fill=teal!16,draw=white,line width=.35pt] (5,-21) rectangle (6,-22);
\node[inner sep=0pt] at (5.5,-21.5) {P};
\filldraw[fill=teal!16,draw=white,line width=.35pt] (6,-21) rectangle (7,-22);
\node[inner sep=0pt] at (6.5,-21.5) {P};
\filldraw[fill=teal!16,draw=white,line width=.35pt] (7,-21) rectangle (8,-22);
\node[inner sep=0pt] at (7.5,-21.5) {P};
\filldraw[fill=teal!16,draw=white,line width=.35pt] (8,-21) rectangle (9,-22);
\node[inner sep=0pt] at (8.5,-21.5) {P};
\filldraw[fill=teal!16,draw=white,line width=.35pt] (9,-21) rectangle (10,-22);
\node[inner sep=0pt] at (9.5,-21.5) {P};
\filldraw[fill=orange!28,draw=white,line width=.35pt] (10,-21) rectangle (11,-22);
\node[inner sep=0pt] at (10.5,-21.5) {F};
\filldraw[fill=teal!16,draw=white,line width=.35pt] (11,-21) rectangle (12,-22);
\node[inner sep=0pt] at (11.5,-21.5) {P};
\filldraw[fill=teal!16,draw=white,line width=.35pt] (12,-21) rectangle (13,-22);
\node[inner sep=0pt] at (12.5,-21.5) {P};
\filldraw[fill=orange!28,draw=white,line width=.35pt] (13,-21) rectangle (14,-22);
\node[inner sep=0pt] at (13.5,-21.5) {F};
\node[anchor=east,rotate=60,inner sep=2pt] at (0.5,-22.15) {Opus 4.8};
\node[anchor=east,rotate=60,inner sep=2pt] at (1.5,-22.15) {K3};
\node[anchor=east,rotate=60,inner sep=2pt] at (2.5,-22.15) {Luna};
\node[anchor=east,rotate=60,inner sep=2pt] at (3.5,-22.15) {Sol};
\node[anchor=east,rotate=60,inner sep=2pt] at (4.5,-22.15) {F5};
\node[anchor=east,rotate=60,inner sep=2pt] at (5.5,-22.15) {F5.1};
\node[anchor=east,rotate=60,inner sep=2pt] at (6.5,-22.15) {Astra};
\node[anchor=east,rotate=60,inner sep=2pt] at (7.5,-22.15) {Qwen};
\node[anchor=east,rotate=60,inner sep=2pt] at (8.5,-22.15) {Terra};
\node[anchor=east,rotate=60,inner sep=2pt] at (9.5,-22.15) {O5.5};
\node[anchor=east,rotate=60,inner sep=2pt] at (10.5,-22.15) {GLM};
\node[anchor=east,rotate=60,inner sep=2pt] at (11.5,-22.15) {V4Pro};
\node[anchor=east,rotate=60,inner sep=2pt] at (12.5,-22.15) {Hy4};
\node[anchor=east,rotate=60,inner sep=2pt] at (13.5,-22.15) {Flash};
\end{tikzpicture}

%% file: grid_table.tex
\begin{table*}[!t]
\centering\fontsize{10}{12}\selectfont\setlength{\tabcolsep}{3.5pt}\renewcommand{\arraystretch}{1.04}
\begin{tabular}{@{}l*{14}{c}@{}}\toprule
Instance & K3 & Luna & Sol & Astra & Terra & Opus & F5 & F5.1 & O5.5 & Qwen & GLM & V4Pro & Hy4 & Flash \\ \midrule
flask-5014 & P & P & P & D & D & P & D & P & P & P & P & D & D & D \\
req-5414 & P & P & P & P & P & P & P & P & P & P & P & P & P & P \\
req-6028 & U & -- & -- & -- & -- & -- & -- & -- & -- & -- & -- & -- & -- & -- \\
xa-3677 & P & P & D & P & P & P & D & P & P & P & P & P & D & D \\
xa-4629 & P & P & P & P & P & P & P & P & P & P & P & P & P & P \\
xa-6721 & P & D & P & P & P & P & P & P & P & D & P & P & P & P \\
pl-4970 & U & -- & -- & -- & -- & -- & -- & -- & -- & -- & -- & -- & -- & -- \\
pl-6903 & P & P & P & P & P & P & P & P & P & P & P & P & P & P \\
pl-7080 & U & -- & -- & -- & -- & -- & -- & -- & -- & -- & -- & -- & -- & -- \\
pt-10051 & X & X & X & X & X & X & X & X & X & X & X & X & X & X \\
pt-10081 & D & D & D & D & D & P & D & D & D & D & D & D & D & D \\
pt-10356 & U & -- & -- & -- & -- & -- & -- & -- & -- & -- & -- & -- & -- & -- \\
pt-5631 & P & P & P & P & P & P & P & P & P & D & P & P & D & D \\
pt-5787 & U & -- & -- & -- & -- & -- & -- & -- & -- & -- & -- & -- & -- & -- \\
pt-5809 & P & P & P & D & D & P & P & P & P & P & P & P & P & P \\
pt-5840 & U & -- & -- & -- & -- & -- & -- & -- & -- & -- & -- & -- & -- & -- \\
pt-6197 & U & -- & -- & -- & -- & -- & -- & -- & -- & -- & -- & -- & -- & -- \\
pt-6202 & P & P & P & D & D & P & P & D & D & D & D & D & D & D \\
pt-7205 & P & P & P & P & P & P & D & P & D & P & D & D & D & D \\
pt-7236 & X & X & X & X & X & X & X & X & X & X & X & X & X & X \\
pt-7324 & U & -- & -- & -- & -- & -- & -- & -- & -- & -- & -- & -- & -- & -- \\
pt-7432 & P & P & P & D & P & P & P & P & D & D & D & P & D & D \\
pt-7490 & D & P & P & D & D & P & D & D & D & D & D & D & D & D \\
pt-7521 & P & P & P & D & D & P & D & D & D & D & D & D & D & D \\
pt-7571 & D & P & P & D & P & P & D & D & D & D & D & D & D & D \\
pt-7982 & P & P & P & D & P & P & D & D & D & D & D & D & D & D \\
pt-8399 & P & P & P & D & D & P & D & D & D & D & D & D & D & D \\
sp-9230 & D & D & D & P & D & D & P & P & D & P & D & D & D & P \\
sp-9591 & P & P & D & D & D & P & P & P & P & P & D & D & P & P \\
sy-12419 & U & -- & -- & -- & -- & -- & -- & -- & -- & -- & -- & -- & -- & -- \\
sy-12481 & P & D & D & P & D & P & P & D & P & D & P & D & D & D \\
sy-12489 & U & -- & -- & -- & -- & -- & -- & -- & -- & -- & -- & -- & -- & -- \\
sy-13091 & U & -- & -- & -- & -- & -- & -- & -- & -- & -- & -- & -- & -- & -- \\
sy-13480 & P & P & P & P & P & P & P & P & P & P & P & P & P & P \\
sy-13551 & P & P & D & P & P & P & P & P & P & P & D & P & P & D \\
\bottomrule
\end{tabular}
\caption{First-round outcome of every Task Preservation cell under the fourteen backbone summarizers (column headers use short forms: K3 = Kimi-K3, Luna = GPT-5.6-Luna, Sol = GPT-5.6-Sol, Terra = GPT-5.6-Terra, Opus = Claude Opus 4.8, F5 = Claude Fable 5, F5.1 = Claude Fable 5.1, O5.5 = Claude Opus 5.5, Hy4 = Hy4-Preview, Astra = GPT-6-Astra, Qwen = Qwen3.8-Max, GLM = GLM-5.3, V4Pro = DeepSeek-V4-Pro, Flash = DeepSeek-Flash). P: preserved (both conditions pass); D: damage (PRE passes, POST fails); R: rescue (PRE fails, POST passes); U: unsolved (both fail); X: excluded PRE-failing instance; --: not evaluated (the cell is outside the cross-backbone PRE-pass comparison).}
\label{tab:full_grid}\end{table*}

%% file: k3_table.tex
\begin{table*}[!t]
\centering
\begin{minipage}[t]{0.32\textwidth}\centering
\fontsize{9}{11}\selectfont\setlength{\tabcolsep}{2pt}
\begin{tabular}{@{}llcccl@{}}\toprule
Model & Instance & $r_1$ & $r_2$ & $r_3$ & Result \\ \midrule
K3 & pt-10081 & D & P & D & SD \\
K3 & pt-7490 & D & P & D & SD \\
K3 & pt-7571 & D & P & D & SD \\
K3 & sp-9230 & D & U & D & SD \\
Luna & xa-6721 & D & D & D & SD \\
Luna & pt-10081 & D & D & D & SD \\
Luna & sp-9230 & D & P & D & SD \\
Luna & sy-12481 & D & D & D & SD \\
Sol & xa-3677 & D & P & D & SD \\
Sol & pt-10081 & D & D & D & SD \\
Sol & sp-9230 & D & D & D & SD \\
Sol & sp-9591 & D & P & D & SD \\
Sol & sy-12481 & D & D & D & SD \\
Sol & sy-13551 & D & D & P & SD \\
Astra & flask-5014 & D & D & D & SD \\
Astra & pt-10081 & D & D & D & SD \\
Astra & pt-5809 & D & P & P & V \\
Astra & pt-6202 & D & D & D & SD \\
Astra & pt-7432 & D & D & D & SD \\
Astra & pt-7490 & D & D & D & SD \\
Astra & pt-7521 & D & D & D & SD \\
Astra & pt-7571 & D & D & D & SD \\
Astra & pt-7982 & D & D & D & SD \\
Astra & pt-8399 & D & D & D & SD \\
Astra & sp-9591 & D & P & P & V \\
Terra & flask-5014 & D & P & D & SD \\
Terra & pt-10081 & D & D & D & SD \\
Terra & pt-5809 & D & P & D & SD \\
Terra & pt-6202 & D & D & D & SD \\
Terra & pt-7490 & D & D & D & SD \\
Terra & pt-7521 & D & D & D & SD \\
Terra & pt-8399 & D & D & D & SD \\
Terra & sp-9230 & D & P & D & SD \\
Terra & sp-9591 & D & D & D & SD \\
Terra & sy-12481 & D & P & D & SD \\
Opus & sp-9230 & D & D & D & SD \\
F5 & flask-5014 & D & D & D & SD \\
F5 & xa-3677 & D & P & P & V \\
F5 & pt-10081 & D & D & D & SD \\
F5 & pt-7205 & D & D & D & SD \\
F5 & pt-7490 & D & D & D & SD \\
F5 & pt-7521 & D & D & D & SD \\
\bottomrule
\end{tabular}
\end{minipage}\hfill
\begin{minipage}[t]{0.32\textwidth}\centering
\fontsize{9}{11}\selectfont\setlength{\tabcolsep}{2pt}
\begin{tabular}{@{}llcccl@{}}\toprule
Model & Instance & $r_1$ & $r_2$ & $r_3$ & Result \\ \midrule
F5 & pt-7571 & D & D & D & SD \\
F5 & pt-7982 & D & D & D & SD \\
F5 & pt-8399 & D & D & D & SD \\
F5.1 & pt-10081 & D & D & D & SD \\
F5.1 & pt-6202 & D & D & P & SD \\
F5.1 & pt-7490 & D & D & D & SD \\
F5.1 & pt-7521 & D & D & D & SD \\
F5.1 & pt-7571 & D & D & D & SD \\
F5.1 & pt-7982 & D & D & D & SD \\
F5.1 & pt-8399 & D & D & D & SD \\
F5.1 & sy-12481 & D & D & D & SD \\
O5.5 & pt-10081 & D & D & D & SD \\
O5.5 & pt-6202 & D & D & P & SD \\
O5.5 & pt-7205 & D & D & D & SD \\
O5.5 & pt-7432 & D & D & D & SD \\
O5.5 & pt-7490 & D & D & D & SD \\
O5.5 & pt-7521 & D & D & D & SD \\
O5.5 & pt-7571 & D & D & D & SD \\
O5.5 & pt-7982 & D & D & D & SD \\
O5.5 & pt-8399 & D & D & D & SD \\
O5.5 & sp-9230 & D & P & D & SD \\
Qwen & xa-6721 & D & P & P & V \\
Qwen & pt-10081 & D & D & D & SD \\
Qwen & pt-5631 & D & D & D & SD \\
Qwen & pt-6202 & D & D & D & SD \\
Qwen & pt-7432 & D & P & D & SD \\
Qwen & pt-7490 & D & D & D & SD \\
Qwen & pt-7521 & D & D & D & SD \\
Qwen & pt-7571 & D & D & D & SD \\
Qwen & pt-7982 & D & D & D & SD \\
Qwen & pt-8399 & D & D & D & SD \\
Qwen & sy-12481 & D & D & P & SD \\
GLM & pt-10081 & D & D & D & SD \\
GLM & pt-6202 & D & D & D & SD \\
GLM & pt-7205 & D & D & P & SD \\
GLM & pt-7432 & D & D & D & SD \\
GLM & pt-7490 & D & D & D & SD \\
GLM & pt-7521 & D & D & D & SD \\
GLM & pt-7571 & D & D & D & SD \\
GLM & pt-7982 & D & D & D & SD \\
GLM & pt-8399 & D & D & D & SD \\
GLM & sp-9230 & D & D & D & SD \\
\bottomrule
\end{tabular}
\end{minipage}\hfill
\begin{minipage}[t]{0.32\textwidth}\centering
\fontsize{9}{11}\selectfont\setlength{\tabcolsep}{2pt}
\begin{tabular}{@{}llcccl@{}}\toprule
Model & Instance & $r_1$ & $r_2$ & $r_3$ & Result \\ \midrule
GLM & sp-9591 & D & P & P & V \\
GLM & sy-13551 & D & D & D & SD \\
V4Pro & flask-5014 & D & D & D & SD \\
V4Pro & pt-10081 & D & D & D & SD \\
V4Pro & pt-6202 & D & D & D & SD \\
V4Pro & pt-7205 & D & D & D & SD \\
V4Pro & pt-7490 & D & D & D & SD \\
V4Pro & pt-7521 & D & D & D & SD \\
V4Pro & pt-7571 & D & D & D & SD \\
V4Pro & pt-7982 & D & D & D & SD \\
V4Pro & pt-8399 & D & D & D & SD \\
V4Pro & sp-9230 & D & P & D & SD \\
V4Pro & sp-9591 & D & P & D & SD \\
V4Pro & sy-12481 & D & D & P & SD \\
Hy4 & flask-5014 & D & D & D & SD \\
Hy4 & xa-3677 & D & D & D & SD \\
Hy4 & pt-10081 & D & D & D & SD \\
Hy4 & pt-5631 & D & D & P & SD \\
Hy4 & pt-6202 & D & D & D & SD \\
Hy4 & pt-7205 & D & D & D & SD \\
Hy4 & pt-7432 & D & D & D & SD \\
Hy4 & pt-7490 & D & D & D & SD \\
Hy4 & pt-7521 & D & D & D & SD \\
Hy4 & pt-7571 & D & D & D & SD \\
Hy4 & pt-7982 & D & D & D & SD \\
Hy4 & pt-8399 & D & D & D & SD \\
Hy4 & sp-9230 & D & P & D & SD \\
Hy4 & sy-12481 & D & P & P & V \\
Flash & flask-5014 & D & D & P & SD \\
Flash & xa-3677 & D & D & P & SD \\
Flash & pt-10081 & D & D & D & SD \\
Flash & pt-5631 & D & D & P & SD \\
Flash & pt-6202 & D & D & D & SD \\
Flash & pt-7205 & D & D & D & SD \\
Flash & pt-7432 & D & D & D & SD \\
Flash & pt-7490 & D & D & D & SD \\
Flash & pt-7521 & D & D & D & SD \\
Flash & pt-7571 & D & D & D & SD \\
Flash & pt-7982 & D & D & D & SD \\
Flash & pt-8399 & D & D & D & SD \\
Flash & sy-12481 & D & D & D & SD \\
Flash & sy-13551 & D & D & D & SD \\
\bottomrule
\end{tabular}
\end{minipage}
\caption{Per-seed verification of first-round damaged Task Preservation cells. P: POST passes; D: POST fails with a passing PRE; U: PRE fails (possible only for Kimi-K3, which reruns both conditions); --: no run. Other backbones resample the summary and POST while reusing the first-seed PRE baseline. SD denotes POST failure in at least two of three seeds; V denotes seed variance and is excluded from stable-damage counts.}
\label{tab:k3}
\end{table*}

%% file: detail_retention_matrix.tex
\begin{table*}[!t]
\centering
\fontsize{10}{12}\selectfont\setlength{\tabcolsep}{5pt}
\begin{tabular}{@{}lccccc@{}}\toprule
Backbone & Claude Code & Codex & Gemini CLI & opencode & Kimi Code \\ \midrule
Kimi-K3 & 15.0\% & 9.5\% & 7.2\% & 9.5\% & 8.2\% \\
GPT-5.6-Luna & 13.3\% & 7.9\% & 6.0\% & 9.1\% & 5.0\% \\
GPT-5.6-Sol & 14.8\% & 8.7\% & 6.1\% & 8.8\% & 7.4\% \\
GPT-5.6-Terra & 8.2\% & \textbf{11.7\%} & 6.5\% & \textbf{12.6\%} & 9.5\% \\
DeepSeek-Flash & \textbf{16.0\%} & 10.7\% & \textbf{7.9\%} & 12.4\% & \textbf{9.7\%} \\
DeepSeek-V4-Pro & 11.9\% & 9.8\% & 6.6\% & 11.2\% & 6.8\% \\
GLM-5.3 & 9.5\% & 8.1\% & 7.8\% & 11.6\% & 6.2\% \\
\bottomrule\end{tabular}
\captionof{table}{Recurring-Detail Retention scores by backbone and prompt (mean per session, $n=144-146$ per cell). Bold marks the best backbone per prompt: DeepSeek-Flash leads on three prompts and GPT-5.6-Terra on two.}
\label{tab:detail-retention-matrix}

\vspace{0.6em}
\begin{minipage}[t]{0.48\textwidth}
\centering\small\setlength{\tabcolsep}{3pt}
\begin{tabular}{@{}lccccc@{}}\toprule
Client & Claude & Codex & Gemini & opencode & Kimi \\ \midrule
Kimi Code & 21.4\% & 13.2\% & 10.1\% & 15.3\% & 11.5\% \\
Claude Code & 9.4\% & 6.4\% & 4.5\% & 3.2\% & 5.5\% \\
opencode & 6.6\% & 3.7\% & 3.8\% & 4.9\% & 4.2\% \\
other & 5.4\% & 2.3\% & 0.6\% & 1.1\% & 0.9\% \\
fake-kimi & 11.8\% & 7.2\% & 8.2\% & 10.6\% & 6.6\% \\
kilocode & 10.9\% & 15.2\% & 8.7\% & 8.7\% & 8.7\% \\
fake-cc & 0.0\% & 0.0\% & 0.0\% & 0.0\% & 0.0\% \\
hertz & 0.0\% & 0.0\% & 0.0\% & 0.0\% & 0.0\% \\
\bottomrule\end{tabular}
\captionof{table}{Recurring-Detail Retention mean score by client and prompt (Kimi-K3; unweighted mean per session). Prompt labels abbreviate Claude Code, Codex, Gemini CLI, opencode, and Kimi Code.}
\label{tab:detail-retention-client-k3}
\end{minipage}\hfill
\begin{minipage}[t]{0.48\textwidth}
\centering\small\setlength{\tabcolsep}{3pt}
\begin{tabular}{@{}lccccc@{}}\toprule
Client & Claude & Codex & Gemini & opencode & Kimi \\ \midrule
Kimi Code & 21.0\% & 13.0\% & 9.2\% & 14.6\% & 8.6\% \\
Claude Code & 6.0\% & 3.2\% & 3.2\% & 3.4\% & 1.3\% \\
opencode & 6.0\% & 1.6\% & 1.2\% & 3.4\% & 0.4\% \\
other & 0.0\% & 0.9\% & 0.9\% & 1.7\% & 3.4\% \\
fake-kimi & 13.1\% & 4.7\% & 2.9\% & 14.5\% & 2.5\% \\
kilocode & 0.0\% & 4.3\% & 6.5\% & 4.3\% & 2.2\% \\
fake-cc & 0.0\% & 0.0\% & 0.0\% & 0.0\% & 0.0\% \\
hertz & 0.0\% & 0.0\% & 0.0\% & 0.0\% & 0.0\% \\
\bottomrule\end{tabular}
\captionof{table}{Recurring-Detail Retention mean score by client and prompt (GPT-5.6-Luna; unweighted mean per session). Prompt labels abbreviate Claude Code, Codex, Gemini CLI, opencode, and Kimi Code.}
\label{tab:detail-retention-client-gpt}
\end{minipage}
\end{table*}

%% file: audit_table.tex
\begin{table}[!htb]
\centering\footnotesize\setlength{\tabcolsep}{3pt}
\begin{tabularx}{\columnwidth}{@{}lcccX@{}}\toprule
Instance & v1 & v2 & v3 & Observation \\ \midrule
sympy-13480 & U & P & P & environment fix; preserved in v3 \\
sympy-13091 & R & U & U & v1 rescue not reproduced \\
xarray-6721 & -- & D & P & v2 path-leak damage \\
requests-5414 & R & -- & P & v1 rescue not reproduced \\
sphinx-9230 & -- & D & D & damage persists in v3 \\
pytest-7521 & -- & D & P & v2 path-leak damage \\
\bottomrule\end{tabularx}
\caption{Interpreter and directory validation (v1: system interpreter; v2: instance interpreter, separate directories; v3: single directory with snapshot reset). Outcome codes as in Table~\ref{tab:full_grid}.}
\label{tab:audit}\end{table}

%% file: instances_table.tex
\begin{table}[!htb]
\centering
\centering\footnotesize\setlength{\tabcolsep}{3pt}
\begin{tabular}{@{}llcl@{}}\toprule
Instance & Repo & \#F2P & Environment \\ \midrule
flask-5014 & flask & 1 & werkzeug<3 \\
req-5414 & requests & 1 & urllib3<2 \\
req-6028 & requests & 2 & urllib3<2 \\
pt-10051 & pytest & 1 & setuptools\_scm shim \\
pt-10081 & pytest & 1 & setuptools\_scm shim \\
pt-10356 & pytest & 1 & setuptools\_scm shim \\
pt-5631 & pytest & 1 & setuptools\_scm shim \\
pt-5787 & pytest & 2 & setuptools\_scm shim \\
pt-5809 & pytest & 1 & setuptools\_scm shim \\
pt-7236 & pytest & 1 & setuptools\_scm shim \\
pt-7324 & pytest & 3 & setuptools\_scm shim \\
pt-7432 & pytest & 1 & setuptools\_scm shim \\
pt-7490 & pytest & 2 & setuptools\_scm shim \\
pt-7571 & pytest & 1 & setuptools\_scm shim \\
pt-8399 & pytest & 1 & setuptools\_scm shim \\
pt-5840 & pytest & 2 & setuptools\_scm shim \\
pt-6197 & pytest & 2 & setuptools\_scm shim \\
pt-6202 & pytest & 1 & setuptools\_scm shim \\
pt-7205 & pytest & 10 & setuptools\_scm shim \\
pt-7521 & pytest & 2 & setuptools\_scm shim \\
pt-7982 & pytest & 1 & setuptools\_scm shim \\
sy-12419 & sympy & 1 & py3.9 (2016-era) \\
sy-12481 & sympy & 1 & py3.9 (2016-era) \\
sy-12489 & sympy & 1 & py3.9 (2016-era) \\
sy-13091 & sympy & 2 & py3.9 (2016-era) \\
sy-13480 & sympy & 1 & py3.9 (2016-era) \\
sy-13551 & sympy & 1 & py3.9 (2016-era) \\
xa-3677 & xarray & 1 & numpy<2, pandas<2 \\
xa-4629 & xarray & 1 & numpy<2, pandas<2 \\
xa-6721 & xarray & 1 & numpy<2, pandas<2 \\
pl-4970 & pylint & 1 & py library for test imports \\
pl-6903 & pylint & 1 & py library for test imports \\
pl-7080 & pylint & 1 & py library for test imports \\
sp-9230 & sphinx & 1 & sphinxcontrib era pins \\
sp-9591 & sphinx & 1 & sphinxcontrib era pins \\
\bottomrule\end{tabular}
\caption{Task Preservation instances with FAIL\_TO\_PASS test counts and environment recipe notes. All environments use pytest$<$8 and a per-instance venv.}
\label{tab:instances}
\end{table}

%% file: variant_grid_table.tex
\begin{table}[!htb]
\centering\footnotesize\setlength{\tabcolsep}{3pt}
\begin{tabular}{@{}lccccc@{}}\toprule
Instance & Claude & Codex & Gemini & opc & Kimi \\ \midrule
flask-5014 & P & P & P & P & P \\
req-5414 & P & F & P & P & F \\
xa-3677 & P & P & P & F & P \\
xa-4629 & P & P & P & P & P \\
xa-6721 & P & P & P & P & P \\
pl-6903 & P & P & P & F & P \\
pt-10051 & X & X & X & X & X \\
pt-10081 & F & F & P & F & P \\
pt-5631 & P & P & P & P & P \\
pt-5809 & P & P & P & P & P \\
pt-6202 & P & P & P & P & P \\
pt-7205 & P & P & P & P & P \\
pt-7236 & X & X & X & X & X \\
pt-7432 & P & P & P & P & P \\
pt-7490 & F & P & P & P & P \\
pt-7521 & P & P & P & F & P \\
pt-7571 & F & F & F & F & P \\
pt-7982 & P & P & P & P & P \\
pt-8399 & P & F & F & P & P \\
sp-9230 & F & F & F & F & F \\
sp-9591 & P & P & P & P & P \\
sy-12481 & P & F & F & P & P \\
sy-13480 & P & P & P & P & P \\
sy-13551 & P & P & F & P & F \\
\bottomrule
\end{tabular}
\caption{First-round Task Preservation outcomes under the five production compaction prompts with Kimi-K3 as the summarizer. Codes: P = POST passes; F = POST fails; X = excluded PRE-failing instance. All non-excluded listed cells pass the PRE control. Prompt headers abbreviate Claude Code, Codex, Gemini CLI, opencode, and Kimi Code.}
\label{tab:variant_grid}\end{table}

%% file: trace_table.tex
\begin{table*}[!tp]
\centering\footnotesize\setlength{\tabcolsep}{3pt}
\begin{tabularx}{\textwidth}{@{}clXX@{}}\toprule
\# & tool & PRE (full history) & POST (summary only) \\ \midrule
1 & \texttt{c}/\texttt{r} & sed -n 290,340p sphinx/\hspace{0pt}domains/\hspace{0pt}python.\hspace{0pt}py & sphinx/\hspace{0pt}util/\hspace{0pt}docfields.\hspace{0pt}py \\
2 & \texttt{c} & cd /\hspace{0pt}tmp \&\& rm -rf reprotest \&\& mkdir reprotest \&\& cd reprotest \&\& cat \textgreater{} index.\hspace{0pt}rst \textless{}\textless{}'\ldots & cd /\hspace{0pt}testbed \&\& git log --oneline -3 \&\& git status \\
3 & \texttt{c} & cd /\hspace{0pt}tmp/\hspace{0pt}reprotest \&\& python - \textless{}\textless{}'EOF' import re html = open('\_build/\hspace{0pt}index.\hspace{0pt}html').\hspace{0pt}read\ldots & pwd \&\& ls \&\& git log --oneline -3 2\textgreater{}/\hspace{0pt}dev/\hspace{0pt}null \textbar{}\hspace{0pt} head \\
4 & \texttt{c} & cd /\hspace{0pt}tmp/\hspace{0pt}reprotest \&\& python - \textless{}\textless{}'EOF' import re html = open('\_build/\hspace{0pt}index.\hspace{0pt}html').\hspace{0pt}read\ldots & sed -n 190,320p sphinx/\hspace{0pt}util/\hspace{0pt}docfields.\hspace{0pt}py \\
5 & \texttt{c} & sed -n 295,310p sphinx/\hspace{0pt}util/\hspace{0pt}docfields.\hspace{0pt}py & grep -n "split\textbackslash{}\textbar{}\hspace{0pt}import\textbackslash{}\textbar{}\hspace{0pt}def " sphinx/\hspace{0pt}util/\hspace{0pt}docfields.\hspace{0pt}py \textbar{}\hspace{0pt} head -40 \\
6 & \texttt{c} & cd /\hspace{0pt}home/\hspace{0pt}user/\hspace{0pt}sphinx 2\textgreater{}/\hspace{0pt}dev/\hspace{0pt}null \textbar{}\hspace{0pt}\textbar{}\hspace{0pt} pwd; python - \textless{}\textless{}'EOF' import re path = 'sphinx/\hspace{0pt}ut\ldots & sed -n 890,1010p tests/\hspace{0pt}test\_domain\_py.\hspace{0pt}py \\
7 & \texttt{c} & sed -n 296,316p sphinx/\hspace{0pt}util/\hspace{0pt}docfields.\hspace{0pt}py & sed -n 840,895p tests/\hspace{0pt}test\_domain\_py.\hspace{0pt}py \\
8 & \texttt{c} & python - \textless{}\textless{}'EOF' path = 'sphinx/\hspace{0pt}util/\hspace{0pt}docfields.\hspace{0pt}py' src = open(path).\hspace{0pt}read() old = """ \ldots & grep -rn "9213\textbackslash{}\textbar{}\hspace{0pt}dict(str" tests/\hspace{0pt}test\_domain\_py.\hspace{0pt}py tests/\hspace{0pt}test\_util\_docfields.\hspace{0pt}py 2\textgreater{}/\hspace{0pt}dev\ldots \\
9 & \texttt{c} &  & grep -n "def test\_docfields" tests/\hspace{0pt}test\_domain\_py.\hspace{0pt}py \\
10 & \texttt{c} &  & sed -n 760,840p tests/\hspace{0pt}test\_domain\_py.\hspace{0pt}py \\
11 & \texttt{c} &  & grep -n "Dict\textbackslash{}[str, str\textbackslash{}] params" tests/\hspace{0pt}test\_domain\_py.\hspace{0pt}py \\
12 & \texttt{c} &  & sed -n 905,935p tests/\hspace{0pt}test\_domain\_py.\hspace{0pt}py \\
13 & \texttt{c} &  & cat \textgreater{} /\hspace{0pt}tmp/\hspace{0pt}repro.\hspace{0pt}py \textless{}\textless{}'EOF' import re from docutils import nodes from sphinx.\hspace{0pt}testing.\ldots \\
\bottomrule\end{tabularx}
\caption{Paired sphinx-9230 action traces ($r_1$, Kimi-K3 summarizer), truncated to 85 characters per action. Tools: c = \texttt{run\_command}, r = \texttt{read\_file}; differing tools are listed as PRE/POST. Rows align action order. PRE ends after 8 actions and POST after 13, within the eight-step continuation budget; the harness judges FAIL\_TO\_PASS tests afterward. PRE applies the repair, while POST continues inspecting tests and reproduction code.}
\label{tab:trace}
\end{table*}

%% file: redo_table.tex
\begin{table}[!tb]
\centering\footnotesize
\begin{tabular}{@{}lcc@{}}\toprule
Cell & PRE redo-reads & POST redo-reads \\ \midrule
pytest-10081 & 1 & 0 \\
pytest-7490 & 1 & 0 \\
pytest-7571 & 0 & 1 \\
sphinx-9230 & 1 & 1 \\
\bottomrule\end{tabular}
\caption{Repeated \texttt{read\_file} calls targeting files read in Phase A for the four stably damaged cells, Kimi-K3 summarizer.}
\label{tab:redo}
\end{table}

%% file: reproducibility_details.tex
\section{Supplementary Configuration and Reference Rules}
\label{app:reprodetails}

\paragraph{Contents and scope.}
The supplementary code implements transcript serialization, structured-value extraction, task labels, retention scoring, and completed-write detection as pure functions. Configuration files specify coding-agent inputs, tool schemas, budgets, and environment recipes; the instance manifest maps repairs to public commits and tests. Production prompts appear in Appendix~\ref{app:prompts}, and artifact availability is specified in Appendix~\ref{app:harness}.

\paragraph{Session filtering.}
Within a session, the sessionizer sorts requests by timestamp and request identifier. Pairs crossing a new compaction marker or a token drop below half the preceding count are excluded from the completeness denominator. Other pairs must have matching nonempty first-200-character message prefixes. When token counts are available, the next cached-token count must be at least 80\% of the preceding total; missing fields do not fail that check. Completeness is the passing fraction, or 1 with no eligible pairs, rounded to three decimals before filtering at 0.8.

\paragraph{Recurring-Detail Retention sampling and windows.}
Section~\ref{sec:data} defines the eligibility filters and boundary index. For requested sample budget $N$, the constructor takes the first $\lfloor N/3\rfloor$ candidates from Claude Code, $\lfloor N/2\rfloor$ from Kimi Code, and $\lfloor N/5\rfloor$ from each other client. Groups follow client-name order and the concatenation is capped at $N$. Missing payloads and samples with fewer than three matching recurring details are then removed.

Recurring-Detail Retention extracts structured values from the first 400{,}000 characters of Python JSON serialization with \texttt{ensure\_ascii=False}. Future text concatenates the first 150{,}000 characters of each available message payload from at most eight requests, without history deduplication. Production-Chain Retention uses the same per-request cap over at most five requests and omits the history intersection.

\paragraph{Recurring-detail extraction and scoring.}
The module specifies path whitelists and exclusions, slash normalization, SHA and other identifier patterns, and URL truncation to 120 characters. Deduplication is within each kind; equal values in different kinds remain distinct labels. Hits use case-sensitive substring membership. Recurring-Detail Retention averages non-null per-session scores without weighting sessions by their number of labels. API/parse errors are excluded; returned empty summaries score zero. The mechanical reference keeps at most 40 sorted values per kind from the bounded history extraction.

\paragraph{Production-Chain Retention disorder and aggregation.}
Successful-write patterns scan the immediately preceding raw message payload. A detected file's basename must occur within 150 characters on either side of a pending phrase in the summary; any match flags the boundary. Missing prior payloads and payloads with no detected writes are recorded as unflagged. The rate therefore measures contradictions captured by these patterns. All retained boundaries enter its denominator; boundaries with at least one evaluable detail enter retention. The last-to-first ratio divides the final retained boundary's retention by the first non-null retention, excluding null final values and zero denominators. Every first boundary in the reported corpus is scorable. The module contains the literal multilingual patterns and summary-extraction rules.

\paragraph{Agent inputs and environment.}
The configuration records the Chinese task/continuation inputs, English POST wrapper, tool schemas, absence of a separate system message, and budgets. Environment recipes specify the pytest version shim, installation order, dependency version constraints, and test command. Original repository and SWE-bench terms apply to the public instances.